\documentclass[twocolumn,aps,pre, reprint,floatfix]{revtex4-1}

\usepackage[english]{babel} 
\usepackage{units}
\usepackage{wrapfig}
\usepackage{amsmath}
\usepackage{url}
\usepackage{hyperref}
\usepackage{amssymb}
\usepackage[utf8]{inputenc}
\usepackage{float}
\usepackage{todonotes}
\usepackage{subcaption}

\newcommand{\pik}{\textsuperscript{1}Potsdam Institute for Climate Impact Research (PIK) - Member of the Leibnitz Association, Telegrafenberg, 14473 Potsdam, Germany}
\newcommand{\hu}{\textsuperscript{2}Department of Physics, Humboldt University, Newtonstraße 15, 12489 Berlin, Germany}
\newcommand{\fub}{\textsuperscript{3}Institute for Meteorology, Free University, Carl-Heinrich-Becker-Weg 6-10, 12165 Berlin, Germany}
\newcommand{\magdeu}{\textsuperscript{4}Department of Water, Environment, Construction and Safety, Magdeburg-Stendal University of Applied Sciences, Breitscheidstra{\ss}e 2, 39114 Magdeburg, Germany}

\makeatletter
\renewcommand{\fnum@figure}{Fig.~\thefigure}
\makeatother

\begin{document}

\title{Edge directionality properties in complex spherical networks}
\author{Frederik Wolf\textsuperscript{1,2}, Catrin Kirsch\textsuperscript{1,3}, Reik V. Donner\textsuperscript{1,4} }
\affiliation{\pik}
\affiliation{\hu}
\affiliation{\fub}
\affiliation{\magdeu}
\date{\today}

\begin{abstract}
Spatially embedded networks have attracted increasing attention in the last decade. In this context, new types of network characteristics have been introduced which explicitly take spatial information into account. Among others, edge directionality properties have recently gained particular interest. In this work, we investigate the applicability of mean edge direction, anisotropy and local mean angle as geometric characteristics in complex spherical networks. By studying these measures, both analytically and numerically, we demonstrate the existence of a systematic bias in spatial networks where individual nodes represent different shares on a spherical surface, and describe a strategy for correcting for this effect. Moreover, we illustrate the application of the mentioned edge directionality properties to different examples of real-world spatial networks in spherical geometry (with or without the geometric correction depending on each specific case), including functional climate networks, transportation and trade networks. In climate networks, our approach highlights relevant patterns like large-scale circulation cells, the El Ni\~{n}o--Southern Oscillation and the Atlantic Ni\~{n}o. In an air transportation network, we are able to characterize distinct air transportation zones, while we confirm the important role of the European Union for the global economy by identifying convergent edge directionality patterns in the world trade network.
\end{abstract}

\maketitle

\section{Introduction}
During the last decades, complex network theory has become a vibrant and growing research field, which utilizes graphs as representations of complex systems \cite{Strogatz2001,Albert2002,Newman2003,Boccaletti2006}. In diverse fields like sociology, neuroscience, and Earth system science, networks have become a powerful tool to investigate interrelations among multiple interacting entities \cite{Dunne2002,Girvan2002,Tsonis2004,Zhou2006,Landherr2010}.

In many real-world complex networks nodes are placed in a metric space and are therefore characterized by a specific location \cite{Gastner2006,Masucci2009, Barthelemy2011}. Hence, an increasing number of studies has discussed the properties of such spatially embedded networks \cite{Bialonski2010,Chan2011, Schultz2014, Wiedermann2017c}. Specifically, different measures have been proposed for complementing topological information by considering the spatial coordinates of nodes \cite{Barthelemy2011}. One recent example of such a measure is the edge anisotropy, which takes the spatial directionality of edges adjacent to a given node  into account. It has been shown that edge anisotropy supplements traditional topological measures like degree or betweenness by indicating the orientation of flows underlying networks constructed from spatio-temporal data \cite{Molkenthin2017}.

In real-world spatially embedded networks, not all nodes may necessarily represent the same area or volume.  Common examples include transportation as well as functional climate and brain networks \cite{Banavar1999,Serrano2003,Achard2006, Donges2009}. In climate networks, for instance, correlations between time series from  differently sized  grid-cells are used to abstract climate variability into a network representation \cite{Tsonis2004,Donges2009,Donner2017}. To respect varying areas of representation (AOR) diverse node-weighted measures have been proposed \cite{Heitzig2012,Wiedermann2013,Zemp2014,Wiedermann2017}.

In this work, we pursue two main objectives. On the one hand, we discuss how to combine edge directionality properties and differing AORs. Specifically, we use analytical considerations together with a numerical toy example to identify biases which occur when investigating edge directionality in spherical networks with heterogeneous AORs. We demonstrate that such biases can be corrected by introducing proper edge weights. On the other hand, we discuss some examples of real-world spherical networks to highlight which type of additional information can be obtained by using edge directionality properties to supplement classical network characteristics. By addressing both aforementioned aspects, we introduce a framework for quantitatively characterizing edge directionality in networks embedded in spherical geometry.

According to this two-fold objective, this paper is structured as follows: Initially, we recall the concept of edge anisotropy \cite{Molkenthin2017} and study the mean edge direction as an additional geometric network measure. Secondly, we investigate edge directionality in angularly regular spherical grids, which are often used in climate data sets, and show that the grid parcellation itself induces generic anisotropy. We also introduce a scheme to avoid the occurring biases, which we utilize to study three functional climate networks. Finally, by investigating two additional real-world examples -- an air transportation network and the world trade network of the year 2009 -- we further demonstrate the broad applicability of edge directionality properties (with or without corrections for spherical geometry effects, depending on the specific case) to a wide range of real-world spherical networks.

\section{Theoretical background}
We consider a network $G$ with $N$ nodes and $E$ edges, the topology of which is  encoded in the binary adjacency matrix $ \mathbf{A} $ whose elements $a_{ij}$ are $1$ if node $i$  is connected to node $j$ and $0$ otherwise. For undirected networks, the adjacency matrix is symmetric \cite{Strogatz2001}. However, many complex systems are better described by directed and weighted networks, such as supply chains where goods are transferred to different consumers in differing quantities. In such cases, the adjacency matrix is replaced by a weight matrix $ \mathbf{W} $ with the edge weights as its entries $w_{ij}$ \cite{Newman2003}. 
In addition to the topological information, in spatially embedded graphs each node $i$ is specified by its spatial position.

\subsection{Anisotropy in Euclidean geometry with homogenous areas of representation}
The concept of edge anisotropy \cite{Molkenthin2017} as a local (per-node) and global characteristic of spatially embedded networks has extended previous studies on node-based distributions of edge directions \cite{Gudmundsson2013,Mohajeri2013,Boers2014a,Rheinwalt2016}. In \cite{Molkenthin2017}, the authors describe the direction of an edge by the $d$-dimensional unit vector $\vec{e}_{mn}$, centered at the position of a node $m$ and pointing towards that of another node $n$. Assigning each connection of a single node $m$ a corresponding unit vector enables us to define the \emph{local anisotropy} of the edges adjacent to $m$ \cite{Molkenthin2017},

\begin{align}\label{unweighted anisotropy}
R_m= \frac{1}{k_m} \left|\left|\sum_{n=1}^K a_{mn} \vec{e}_{mn} \right|\right|,
\end{align}
where $\sum_n a_{mn}=k_m$ denotes the degree of node $m$. In the following, we will also refer to the \emph{mean edge direction} \cite{Molkenthin2017},

\begin{align}\label{mean edge direction}
\vec{r}_m= \frac{1}{k_m} \sum_{n=1}^K a_{mn} \vec{e}_{mn}
\end{align}
\noindent with $R_m= ||\vec{r}_m||$.

In weighted networks, we replace the adjacency matrix by the weight matrix and redefine the normalization by the sum of all weights \cite{Molkenthin2017} (known as the node strength) to compute the \emph{weighted local anisotropy} as

\begin{align}\label{weighted anisotropy}
R_m^w= \frac{1}{\sum_n w_{mn}} \left|\left|\sum_{n=1}^K w_{mn} \vec{e}_{mn} \right|\right|.
\end{align}

The definition of the \emph{weighted mean edge direction} follows accordingly.

To account for directed networks, we consider the in- and out-degree. Molkenthin et al.~\cite{Molkenthin2017} defined the in- and out-edge anisotropy as 

\begin{align}
R_m^{in}= \frac{1}{\sum_n a_{mn}} \left|\left|\sum_{n=1}^K a_{mn} \vec{e}_{mn} \right|\right|,\\
R_m^{out}= \frac{1}{\sum_n a_{nm}} \left|\left|\sum_{n=1}^K a_{nm} \vec{e}_{nm} \right|\right|.
\end{align}

For weighted, directed graphs we can again substitute the adjacency matrix with the weight matrix \cite{Molkenthin2017}.

\subsection{Anisotropy in spherical geometry with homogeneous areas of representation}

On spherical surfaces embedded in three-dimensional space (like in good approximation the surface of the Earth), each node $m$ is characterized, in addition to its topological attributes,  by a latitude $\phi_m$ and longitude $\lambda_m$. Using such spherical coordinates we can assign course angles $\gamma_{mn}$ to each edge pointing from node $m$ to node $n$ (for a detailed derivation, see the Appendix) by setting

\begin{align}
\gamma_{mn}=\arccos c_{mn}
\label{eq:courseangle}
\end{align}

with

\begin{align}
c_{mn}=\frac{\cos\phi_m\sin\phi_n-\cos(\lambda_m-\lambda_n)\cos\phi_n \sin\phi_m}{\sqrt{1-\left(\cos(\lambda_m-\lambda_n)\cos\phi_m\cos\phi_n+\sin\phi_m\sin\phi_n\right)^2}}.
\end{align}

Course angles $\gamma_{mn}$ have their origin in navigation and denote the angle between the geodetic north and the course of a moving ship, which we replace here by the edge direction. Since $\gamma_{mn} \in [0,\pi]$, we associate edges with an \emph{eastward} component with the angle $\beta_{mn}=\gamma_{mn}$ and such with an \emph{westward} component with the angle $\beta_{mn}=2\pi-\gamma_{mn}$.

For analyzing a node's edge anisotropy we locally neglect curvature and use polar coordinates to define unit vectors for each edge as $\vec{e}_{mn}=\begin{pmatrix} \sin\beta_{mn}\\ \cos\beta_{mn} \end{pmatrix}$ (with the second coordinate corresponding to the northward component)  which enables us to apply Eq.~(\ref{unweighted anisotropy}).

We call the angle $\delta_m$ which is measured between the mean edge direction $\vec{r}_m$ (Eq.~\ref{mean edge direction}) and the arbitrary reference axis (e.g., the geodetic north) the \emph{local mean angle} of the edges adjacent to node $m$.

In contrast to Euclidean angles, course angles are not as easily determined. In particular, course angles are not symmetric. Up to now, we have considered all links connecting node $m$ with $k_m$ other nodes and studied the angles between the edges and an arbitrary axis centered at node $m$. In turn, we can also account for the angles measured at all other $k_m$ nodes and define the \emph{external anisotropy} as

\begin{align}
R_m^*= \frac{1}{k_m} \left|\left|\sum_{n=1}^K a_{mn} \vec{e}_{nm} \right|\right|
\end{align}

and the \emph{mean external edge direction} as

\begin{align}
\vec{r}_m^*=\frac{1}{k_m}\sum_{n=1}^K a_{mn}\vec{e}_{nm}.
\end{align}

In Euclidean geometry, straightforward calculations show that $R_m^*=R_m$ and $\vec{r}_m=-\vec{r}_m^*$, which, however, does not generally hold for spherical geometry.

\subsection{Anisotropy in spherical geometry with heterogeneous areas of representation}
Heterogenous placement of nodes can induce biases in various network measures. In climate networks, for instance, the consideration of nodes on an \emph{angularly regular} spherical grid (i.e., with constant differences in latitude and longitude between neighboring grid points) systematically biases the resulting network properties due to an overemphasis of nodes close to the poles \cite{Heitzig2012,Radebach2013,Wiedermann2013}. To account for this problem, the framework of \emph{node-splitting invariance} (n.s.i.) has been developed, which makes use of appropriate node weights to correct for such biases \cite{Heitzig2012,Wiedermann2013,Zemp2014}. Specifically, the authors of \cite{Heitzig2012} developed a comprehensive framework to tackle heterogeneous node placement by node splitting and twin merging.  

In the following, we will address the corresponding generic bias of edge directionality properties in a spherical grid. For this purpose, we utilize edge weights, which, in this case, can also be derived from the n.s.i. framework as node weights of the adjacent nodes. However, utilizing node weights could induce ambiguous edge weight allocations, especially for the external anisotropy $R_m^*$. Therefore, we refer to the correction of biases resulting from the heterogeneous area of representation by using the term \emph{AOR correction}.
Next, we propose a general definition of the AOR corrected edge anisotropy before defining proper edge weights for the special case of the local anisotropy in an angularly regular spherical grid. 

We compute the \emph{AOR corrected local anisotropy} by utilizing Eq.~(\ref{weighted anisotropy}) and replacing the edge weights by the AOR weights:
\begin{align}\label{AOR anisotropy}
R_m^{AOR}= \frac{1}{\sum_n w_{mn}^{AOR}} \left|\left|\sum_{n=1}^K w_{mn}^{AOR} \vec{e}_{mn} \right|\right|.
\end{align}

For weighted graphs, we suggest multiplicative combinations between intrinsic weights and the AOR correction. The \emph{AOR corrected weighted local anisotropy} then reads
\begin{align}
R_m^{w,AOR}= \frac{1}{\sum_n w_{mn}^{AOR} w_{mn}} \left|\left|\sum_{n=1}^K w_{mn}^{AOR} w_{mn} \vec{e}_{mn} \right|\right|.
\end{align}

For the external anisotropy, we accordingly suggest
\begin{align}
R_m^{*AOR}= \frac{1}{\sum_n w_{nm}^{AOR}} \left|\left|\sum_{n=1}^K w_{mn}^{AOR} \vec{e}_{nm} \right|\right|.
\end{align}

In the remainder of this work, we will focus on the local anisotropy on homogeneous and heterogenous spherical grids and will not further discuss the external anisotropy.

\subsection{Anisotropy bias on an angularly regular spherical grid}
As already mentioned above, in functional climate networks nodes are often associated with an angularly regular spherical grid with constant latitudinal and longitudinal distances between neighboring grid points. To illustrate possible anisotropy biases, which arise due to the spatial embedding using such grids, we first consider the resulting properties of some synthetic network. In line with previous work \cite{Heitzig2012}, we suggest a rotationally and translationally symmetric network by independently linking nodes on some angularly regular spherical grid depending on their angular distance $\alpha_{mn}$ with the probability 

\begin{align}
p(\alpha_{mn})=\min(1,\exp(0.4-0.09\alpha_{mn})).
\end{align}
\noindent The parameters are chosen such as to result in a link density of approximately $0.035$, which is of the order of values commonly used in climate network studies \cite{Heitzig2012,Donner2017}. For our analysis, we use a single realization of the corresponding random graph on an angularly regular spherical grid with a spatial resolution of approximately $1.875^\circ \times 1.875^\circ$ resulting in $N=18,432$ grid points. The obtained results are extremely stable among different realizations due to the high number of nodes considered (not shown).

\begin{figure*}[t]
    \includegraphics{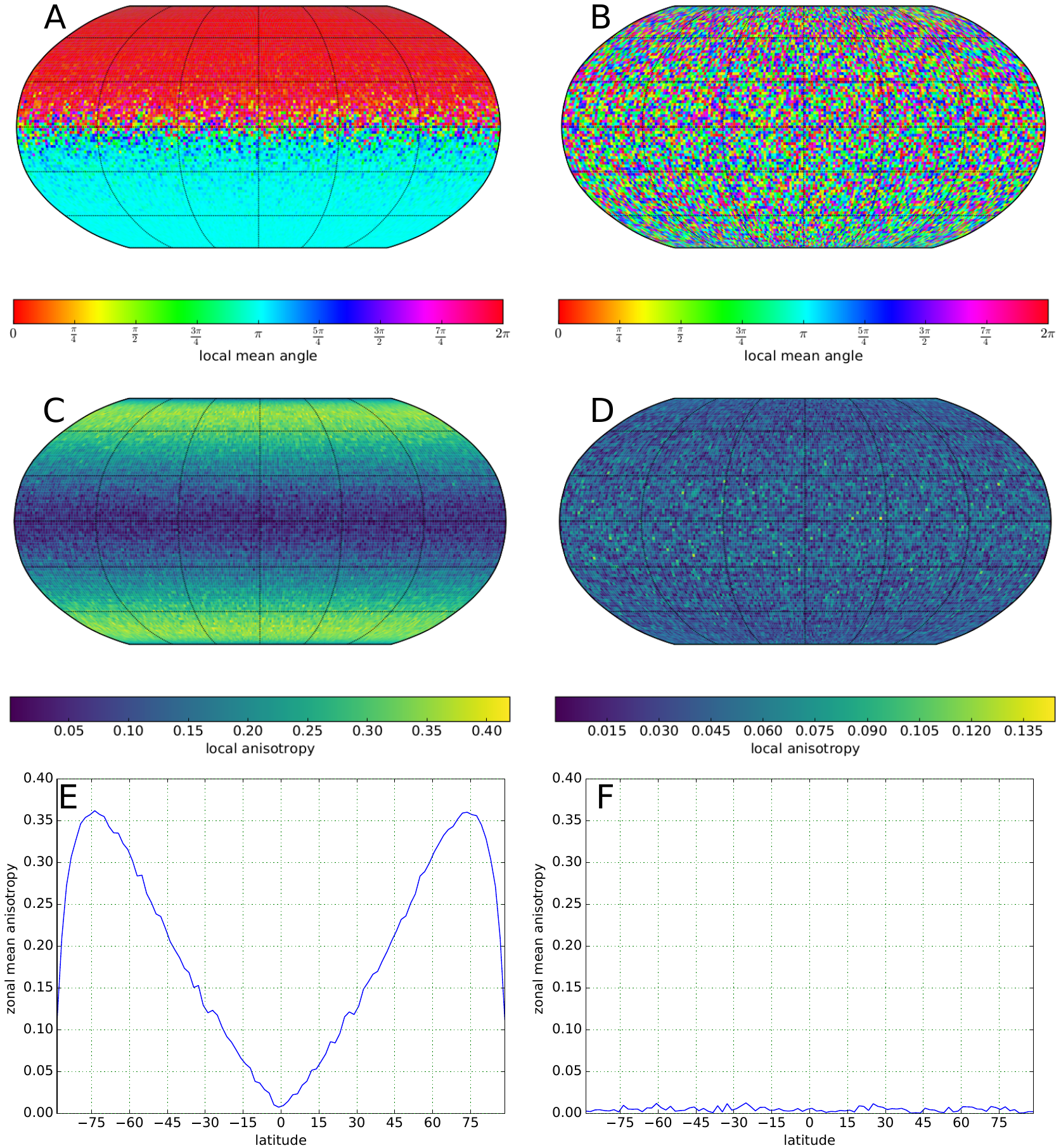}
    \caption{(A,C,E) Uncorrected and (B,D,F) AOR corrected values of (A,B) local mean angle, (C,D) local anisotropy and (E,F) zonal mean anisotropy for one realization  of the benchmark network. Note the different color scales in panels (C) and (D).}\label{figure1}
\end{figure*}

Although we connect nodes independently at random, the synthetic network exhibits nontrivial local anisotropy and local mean angle patterns if not corrected for the bias discussed above, as it is shown in Fig.~\ref{figure1}. Figure~\ref{figure1}A displays the mean angle between geodetic north and the actual mean edge direction. One can distinguish the hemispheres as local mean angles on the northern hemisphere exhibit values around $0$ or $2\pi$ and those in the southern hemisphere around $\pi$. The local anisotropy (Fig.~\ref{figure1}C) increases gradually from minimum values around the equator to higher latitudes and decreases around the poles, which is clearly visible in the corresponding zonal mean (Fig.~\ref{figure1}E). 

The observed behavior apparently contradicts the homogeneous linkage mechanism of our model. The reason for this latitude-dependent bias is therefore not the benchmark network itself, but solely its spatial embedding on the angularly regular spherical grid. Over a vast range of latitudes, only excluding the polar regions, the local anisotropy increases as the spatial node density rises. This heterogeneous node density is a well-known issue on such spherical grids \cite{Heitzig2012}. In particular, the node density increases with the cosine of the latitude, which can be derived in a straightforward manner. Each node is associated with a specific AOR that is proportional to the area $A_\Box$ of the spherical rectangle in which the node is placed. We denote the separation between two neighboring latitudes $\phi_i$ and $\phi_j$ of the considered grid as $\Delta \phi$. This enables us to compute the difference between the areas $A_{1,2}$ of two spherical caps cut at latitudes $\phi_i-\frac{1}{2}\Delta \phi$ and $\phi_i+\frac{1}{2} \Delta \phi$, which in turn is proportional to $A_\Box$. Thus, we can simplify

\begin{align}
\begin{split}
A_{\Box}& \propto A_1-A_2\propto \sin\left(\phi_i+\frac{1}{2}\Delta \phi\right)-\sin\left(\phi_i-\frac{1}{2} \Delta \phi\right)\\
&= 2\sin\frac{\Delta \phi}{2}\cos\phi_i\propto \cos\phi_i.
\end{split}
\end{align}

Let us assume that we randomly choose a node in the northern hemisphere. We observe, that within the vicinity of the node the distance between neighboring nodes differs in different directions. Therefore a distance based linkage mechanism induces more edges that point northward (respectively southward for nodes in the southern hemisphere). The marked reduction at the poles is due to cross-polar edges, which occur when a critical portion of the links traverse the polar region and connect to nodes of a similar latitude but opposite longitude. In our parameter setting, the probability $p(\alpha_{mn})$ decreases to $\frac{1}{e}$ at $\alpha_{mn} \sim 15^\circ$, causing the reduction for latitudes $ \gtrsim |\pm 75^\circ|$.

\subsection{Bias correction scheme for spherical grids}
As shown above, even an angularly regular spherical grid induces a systematic bias in the estimated edge directionality properties. To correct for this bias, we can apply the edge-weighted anisotropy with a proper choice of edge weights. In climate networks, n.s.i. node weights that are chosen proportional to the AOR of a node have already been demonstrated to correct for corresponding biases in topological network characteristics \cite{Heitzig2012,Zemp2014}. Following the same rationale, we propose here to utilize Eq.~(\ref{weighted anisotropy}) and weigh an edge between node $m$ and node $n$ with $w_{mn}^{AOR}=\cos \phi_n$ to obtain unbiased estimates of edge directionality properties like the local anisotropy of node $m$, which is very similar to the correction scheme suggested in the n.s.i. framework \cite{Heitzig2012}. The result of the application of these edge weights is presented in Fig.~\ref{figure1}. Figure~\ref{figure1}B shows the thus corrected mean angles, which exhibit a uniform distribution on $[0,2\pi]$ without spatial pattern. We also do not observe any significant local anisotropy (Fig.~\ref{figure1}D). In our parameter setting, the maximum zonal average anisotropy is approximately by a factor $10^2$ smaller than without the edge weights (Fig.~\ref{figure1}F).

\section{Data}
To further illustrate the potential of edge directionality properties, we consider various data sets from different backgrounds. Initially, we analyze different functional climate networks constructed from aquaplanet simulations and reanalysis data, respectively. However, the applicability of edge directionality properties is not limited to climate networks. To illustrate the general usefulness of such measures, we also discuss some of the corresponding characteristics of an air transportation network and  the world trade network.

\subsection{Functional climate networks}
The construction of climate networks follows an established procedure \cite{Tsonis2006, Donner2017}.  We start with a set of time series at grid points that are located on an angularly regular spherical grid over the Earth's surface. After removing the seasonal cycle by subtracting the annual mean climatology for each grid point, we compute the absolute value of the Pearson correlation coefficient at zero time lag for each pair of time series describing the variability of the observable of interest at the respective grid points. We threshold the entries of the resulting correlation matrix at some value to obtain the adjacency matrix of an associated network representation of the strongest co-variability structures from the underlying spatio-temporal data set. The threshold is set according to the desired link density. A link density of $\rho\approx 0.005$ has been shown to be appropriate for climate networks with a high spatial resolution \cite{Donges2009, Donner2017} and will be employed here. We delete all self-loops as each time series is perfectly correlated with itself at zero lag.

For constructing functional climate networks, we utilize two different data sets.

On one hand, we study a model dataset with aquaplanet simulations performed within the TRACMIP coordinated experiment, where an idealized planet is studied that consists of a thermodynamic slab ocean with interactive sea-surface temperatures and air-sea coupling \cite{Voigt2016,Stevens2013}. 
In particular, we consider surface air temperature (SAT) time series from the ECHAM6.1 AquaControl simulation without continental landmasses, which covers $30$ years of climate dynamics with monthly values and a spatial resolution of approximately $1.875^\circ \times 1.875^\circ$. The simulation utilizes present day insolation and hemispherically asymmetric, but zonally symmetric meridional heat transport resulting in asymmetric circulation cells and a northward shift of the mean inter-tropical convergence zone (ITCZ, mean location at $4.6^\circ \text{N}$) \cite{Voigt2016}.

On the other hand, we utilize global data from the ERA-Interim reanalysis \cite{Dee2011}, which provides continuously updated data of different variables at a resolution  of $2.5^\circ \times 2.5^\circ$.
For our study, we consider monthly precipitation sums and  SAT averages from 1979 to 2016.

\subsection{Air transportation network}
We utilize the \emph{OpenFlights} database (\url{https://openflights.org/data.html}, date: 2 May 2018). In our study, we consider airport and route data sets. The network contains $N=3,425$ airports with locations and $E=37,595$ weighted, directed edges representing served routes, corresponding to an edge density of about $0.3\%$. The weights are chosen according to the number of flights that serve a specific, directed connection.

\subsection{World trade network}
We finally make use of the year 2009 world trade network inferred from the Eora multi-regional input-output database \cite{Lenzen2012}. The network contains $E=7,043$ directed and weighted edges which represent the trade of goods between pairs of $N=186$ countries (with an edge density of about $20\%$). Each node represents a country and its location is defined by its geographical midpoint, while the directed trade volume defines the edge weights.

\section{Results}
In this paper, we aim to outline the broad applicability of edge directionality properties in spherical networks by extracting several meaningful insights from each data set. Therefore, this illustration does not provide a full analysis of the data sets but highlights the applicability of our approach to a wide class of systems.

\subsection{Functional climate networks}

\subsubsection{Aquaplanet SAT climate network}

\begin{figure*}[t]
    \includegraphics{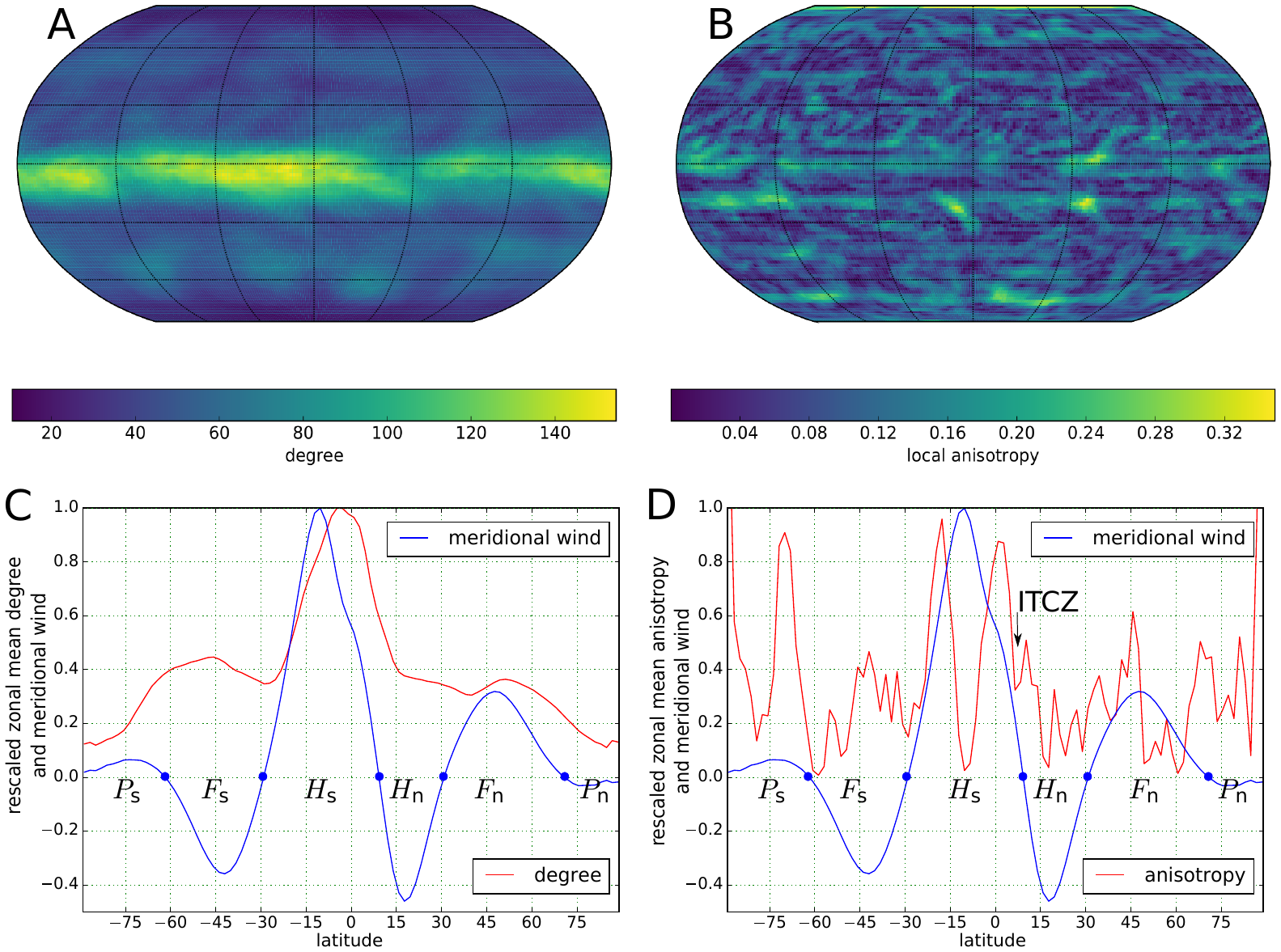}
    \caption{(A) N.s.i. degree and (B) AOR corrected anisotropy of the aquaplanet SAT climate network. (C,D) Corresponding zonal means of (A,B) together with the meridonal wind at 925 hPa (rescaled to dimensionless units) to indicate the circulation cells. The edges of the Hadley ($H_{s/n}$), Ferrel ($F_{s/n}$) and Polar cells ($F_{s/n}$) are indicated, with subscripts specifying the hemisphere.} \label{figure2}
\end{figure*}

Due to the implemented heat flux correction in the aquaplanet simulation, the annual mean ITCZ is shifted northwards causing enlarged circulation patterns on the southern hemisphere as opposed to narrowed northern hemisphere counterparts \cite{Voigt2016}. Specifically, Fig.~\ref{figure2}A,C features a region of elevated degree in the northernmost part of the southern hemisphere Hadley cell, while there is no clear indication of the circulation cells themselves. However, we observe in general a strong similarity between the zonal mean profiles of meridional wind and node degree (Fig.~\ref{figure2}C). In line with previous studies \cite{Molkenthin2017}, we hypothesize that this agreement is caused by fast north/southward directed winds inducing large-scale transport of temperature fluctuations. Other topological network  measures like betweenness or closeness of the SAT climate network either exhibit no directly interpretable patterns or solely indicate the hemispheric asymmetry of the aquaplanet's climate (not shown). 

In comparison to the degree, the resulting local AOR weighted anisotropy reveals more complex spatial patterns, as shown in Fig.~\ref{figure2}B,D. Notably, we observe that at the zonal average, the dominant wind directions are related to the anisotropy and the mean edge direction (not shown) of the SAT network in some non-trivial way, as we will detail in the following.

One interesting observation is that the ITCZ is characterized by a small zonal low anisotropy band, centered north of the equator in line with the mean position of the ITCZ at approximately $4.6^\circ\text{N}$ \cite{Voigt2016}. As indicated by the term \emph{convergence} zone, connections point north- and southward representing oppositely directed near-surface wind patterns and, hence, strong correlations in the SAT fields \cite{Molkenthin2017} (here in meridional direction), resulting in relatively small anisotropy values. Moreover, we recognize the edges of the circulation cells in the southern hemisphere (in terms of vanishing zonal mean meridional wind), which mostly coincide with relatively low zonal mean anisotropy (Fig.~\ref{figure2}D). 

The southern hemisphere Hadley cell ($H_s$) ranges from $\sim 4^\circ\text{N}$ to $\sim 30^\circ\text{S}$ and the Ferrel cell ($F_s$) extends up to $\sim 60^\circ\text{S}$ (Fig.~\ref{figure2}D). Interestingly, we recognize the southern hemisphere Hadley cell as being represented by a double peak in the zonal mean anisotropy. We hypothesize that this could be characteristic for a cross-section (which is here realized by the zonal averaging) through large areas with densely connected nodes as we do not observe any preferred link direction in the central parts, but recognize a preferred link direction towards the interior (not shown) at the outer parts of such regions, followed by a decrease of edge anisotropy at the boundaries, due to a complete change in the preferred edge direction.  

In the northern hemisphere, especially the Hadley cell ($H_n$) is less evident due to the weaker tropical wind fields and smaller cell size. But we clearly see trade wind structures (Southwest-to-Northeast and Northwest-to-Southeast high anisotropy bands on the northern and southern hemisphere, respectively) in Fig.~\ref{figure2}B blurring the edges of the Ferrel ($F_n$) and Hadley cell ($H_n$).

As the most prominent common feature of the six circulation cells, we find some pronounced anisotropy minima in the centers of these cells (with the southern hemisphere Ferrel cell as a possible exception, cf.~Fig.~\ref{figure2}D). To understand this general observation, note again that the edges of these cells are characterized by -- on average -- either convergent or divergent near-surface winds, which lead to some preferred wind direction at the corresponding latitudes. The resulting near-surface wind patterns determine the spatial directions along which fluctuations of temperature are transported (over possibly large distances) with the atmospheric flow \cite{Molkenthin2017}. However, due to seasonal (but possibly also inter-annual) variations in the exact positions of the different circulation cells, the edges of these cells vary and, hence, cannot be identified well by the observed mean anisotropy patterns. Together with previous observations for simple spatio-temporal model systems suggesting that regions of fast directed flow commonly coincide with large degree rather than large anisotropy \cite{Molkenthin2017}, this could explain why the largest anisotropy values are somewhat meridionally shifted with respect to the (mean) boundaries of the circulation cells, whereas marked anisotropy minima emerge in the centers of these cells. We emphasize that even without the variability in the positions of the circulation cells, we would not expect a one-by-one correspondence between the meridional wind strength and the anisotropy patterns because of the existence of multiple superposed effects, like the action of the Coriolis force or possible atmospheric instabilities leading to the formation of large-scale planetary wave patterns, which additionally blur the general pattern of the circulation cells as suggested by the marked spatial variability of edge anisotropy along most latitudinal bands (Fig.~\ref{figure2}B).

\subsubsection{Real-world climate networks - SAT field}

\begin{figure*}[t]
    \includegraphics{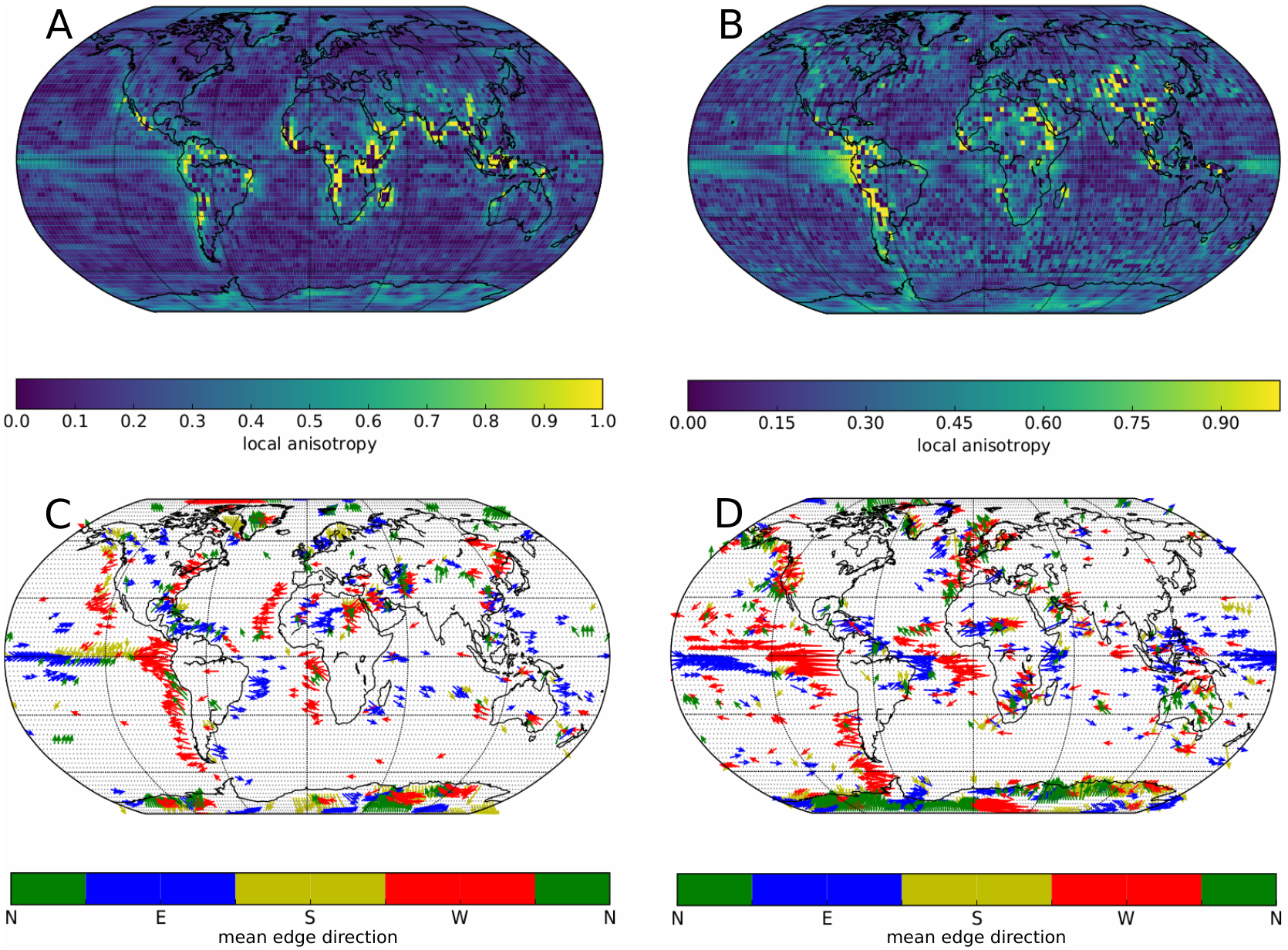}
  \caption{(A,B) Local anisotropy and (C,D) mean edge direction of the (A,C) SAT and (B,D) precipitation network from the ERA-Interim data for all nodes with $R_m>0.25$ and $k_m>10$. The color code indicates the cardinal direction. Areas with degree zero are assigned zero anisotropy.} \label{figure3}
\end{figure*}

Figure~\ref{figure3}A,C shows the AOR corrected local anisotropy values and the mean edge directions for the climate network representation of the real-world SAT field, highlighting the continental coastlines by elevated values and markedly directed structures, respectively. To understand this observation, note that a first-order approximation of the global temperature variability would distinguish between land and ocean regions as the heat capacity of large water masses dampens temperature fluctuations. Hence, the coastal regions are characterized by high anisotropy values and coincide with a localized change of the preferred edge directionality. Especially the west coasts appear to be highly anisotropic zones, as the zonal west winds get deflected by orography and pressure patterns in coastal areas. Additionally, we observe lower degree over land masses possibly reflecting the larger persistence of SAT over the oceans leading to spatial auto-correlations and, hence, a tendency towards a higher number of connections (not shown). We therefore mainly show the mean edge direction over the ocean in coastal areas in Fig.~\ref{figure3}C. 

The El Ni\~{n}o--Southern Oscillation (ENSO) region in the tropical eastern-to-central Pacific also appears highlighted in Fig.~\ref{figure3}A and Fig.~\ref{figure3}C, because the ENSO exhibits highly correlated temperature variations on a monthly scale, which induces elevated degree in the corresponding region.

\subsubsection{Real-world climate networks - precipitation}
We also constructed climate networks from the ERA-Interim precipitation data. We are aware, that there are known issues with the cloud cover in the ERA-Interim analysis \cite{ERAINT}, but, in contrast to local rainfall events, large-scale precipitation patterns at monthly time scales can still be reproduced with sufficient accuracy for the purpose of this study. 

Figure~\ref{figure3}D shows the mean edge directions of the precipitation network, which reveal several features. Firstly, we again observe an ENSO related structure in the eastern tropical Pacific. Similarly to the double band structure of the anisotropy in the previously described aquaplanet scenario (Fig.~\ref{figure2}D), which had represented the southern hemisphere Hadley cell, here  the edge direction exhibits two regions of opposite orientation, with a transition region in between, representing the ENSO (Fig.~\ref{figure3}D). Secondly, we also identify a clear, but less pronounced pattern spatially coinciding with the Atlantic Ni\~{n}o region between South America and Africa in Fig.~\ref{figure3}D \cite{Keenlyside2007}. This is remarkable as we did not analyze the topology of the network but only considered the spatial distribution and orientation of links.

\subsection{Air transportation network}
Airline networks have previously been analyzed revealing complex degree distributions and topologies \cite{Vespignani2004, Gastner2006}. Here we study a weighted and directed air transportation network as an example for the application of edge directionality properties to  transportation networks.

\begin{figure}
    \includegraphics[width=\linewidth]{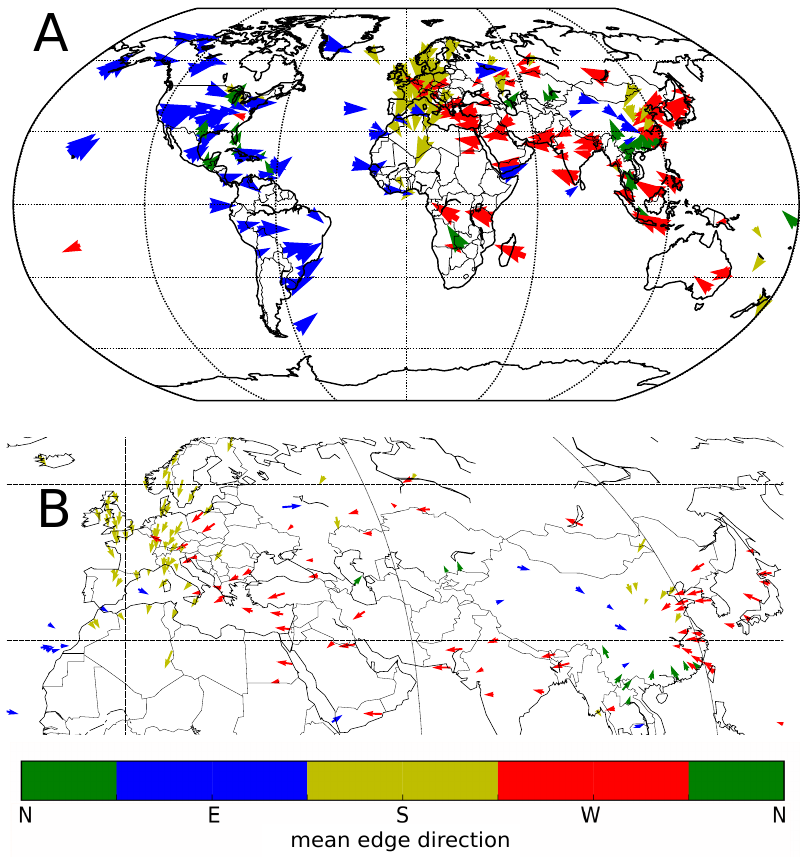}
  \caption{(A) Mean edge direction of the air transportation network for all nodes with $R_m>0.3$ and $k_m>25$. (B) Mean edge direction of the air transportation network over Eurasia. The color code indicates the cardinal direction.} \label{flights_fig}
\end{figure}

Figure~\ref{flights_fig} shows the mean edge directions of the outgoing flights, for the approximately $600$ of $7,184$ airports with $R_m>0.3$ and $k_m>30$, colored according to their cardinal direction. Thereby, we filter out all weakly anisotropic nodes and nodes with just a few connections. Note that we do not use any AOR correction in this example. Here, nodes directly represent the airports rather than large-scale regions of possibly different size, which is why there is no AOR related bias in the network structure that needs to be corrected for. However, airports obviously exhibit different sizes and heterogeneous spatial locations, so that including additional AOR weights may provide interesting complementary information. Since we only attempt here to highlight the general applicability of edge directionality properties, we refrain from a more detailed analysis of the resulting properties.

Figure~\ref{flights_fig}A reveals several features of the global air transportation network. Only a minority of all outgoing flights cross the Pacific ocean: All airports along the North and South American west coast exhibit large anisotropy towards the east, while Asia's coastal airports have predominately westward pointing outgoing connections. Apparently, in the air transportation network, we cannot argue, that coastal airports mainly serve short distance flights. If this held, all coastal airports would exhibit mean edge directions pointing toward the continental interiors. In turn, the majority of the  North and South American east coast airports also exhibit mean edge directions pointing eastward crossing the Atlantic ocean. 

The Eurasian air traffic exhibits two additional interesting features (Fig.~\ref{flights_fig}B):
Firstly, most highly anisotropic nodes in Europe show mean edge directions pointing southward. This is due to the fact that most important global destinations are located further south than Europe while zonal (East/West) components average out.
Secondly, East Asia shows a very interesting and different edge directionality pattern. The airspace over central China is the hotspot of the East Asian air transportation. Most surrounding airports (in Korea, Japan, South East Asia, Mongolia, western China) exhibit mean edge directions pointing towards or across this region.

\subsection{World trade network}
Trade networks have been vastly studied with different types of complex network approaches \cite{Serrano2003, Maluck2015}. In the context of this work, we consider the global trade network from 2009 \cite{Lenzen2012}.

\begin{figure}
    \includegraphics[width=\linewidth]{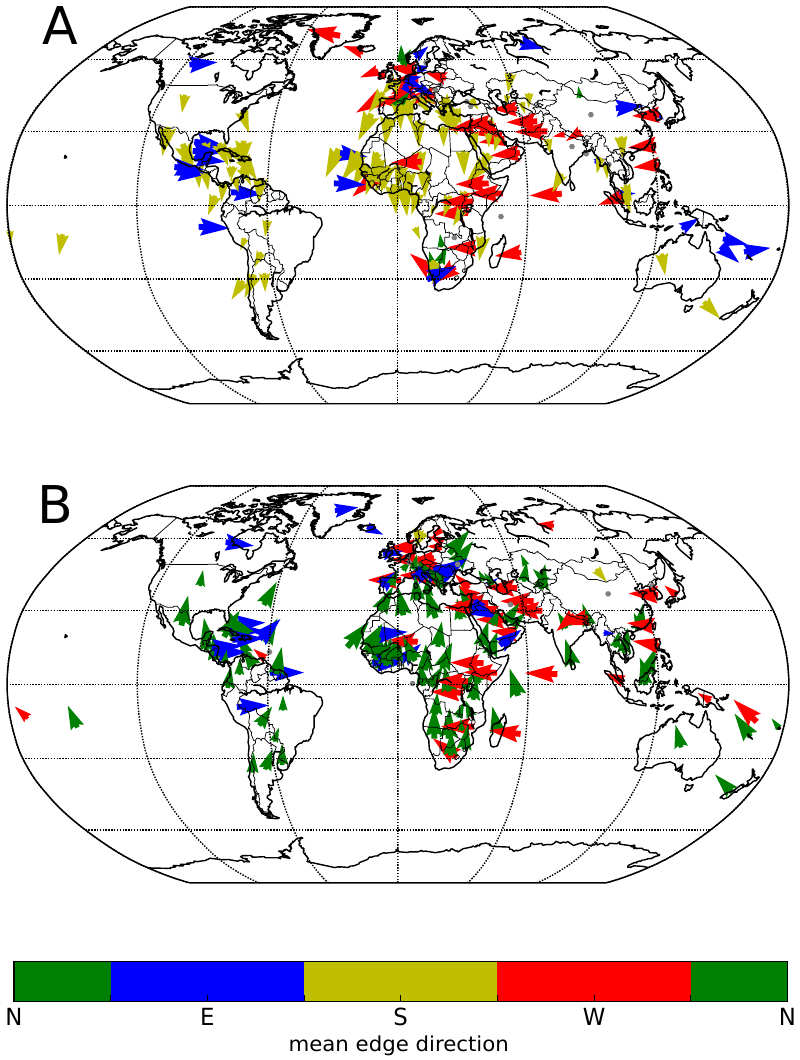}
  \caption{Mean edge direction of the global (A) import and (B) export network of 2009 for all nodes with $R_m>0.3$ and $k_m>25$. The color code indicates the cardinal direction. The arrows point towards (import) or away from (export) the centers of the respective countries. }\label{world trade}
\end{figure}

Figure~\ref{world trade} shows the mean edge directions of the weighted and directed world trade network for both, the incoming (import, Fig.~\ref{world trade}A) and outgoing flow of products and services (export, Fig.~\ref{world trade}B). Again, we threshold the anisotropy at $R_m=0.3$ and the degree at $k_m=25$, which removes only $15$ of $186$ nodes. The world trade network is highly anisotropic and densely connected as only a few countries do not exceed these thresholds. In line with the previous example, we do not employ a correction for the AOR since individual economic relevance has already been accounted for in the edge weights representing the monetary value of mutual demand and supply relationships.

One striking feature is that import and export patterns can be clearly  distinguished. The difference is an inherent aspect of trade and heterogeneous among countries. On the one hand, most countries exhibit opposite import and export directions. On the other hand, there are countries like Canada for which the import and export directions align (Canada apparently receives a vast amount of imports from Asian counties and exports mainly eastward). One reason for this heterogeneity is that some countries represent intermediate elements of global supply chains, while others rather combine different goods to complete high-end technical devices. 

\begin{figure}
\includegraphics[width=0.48\textwidth]{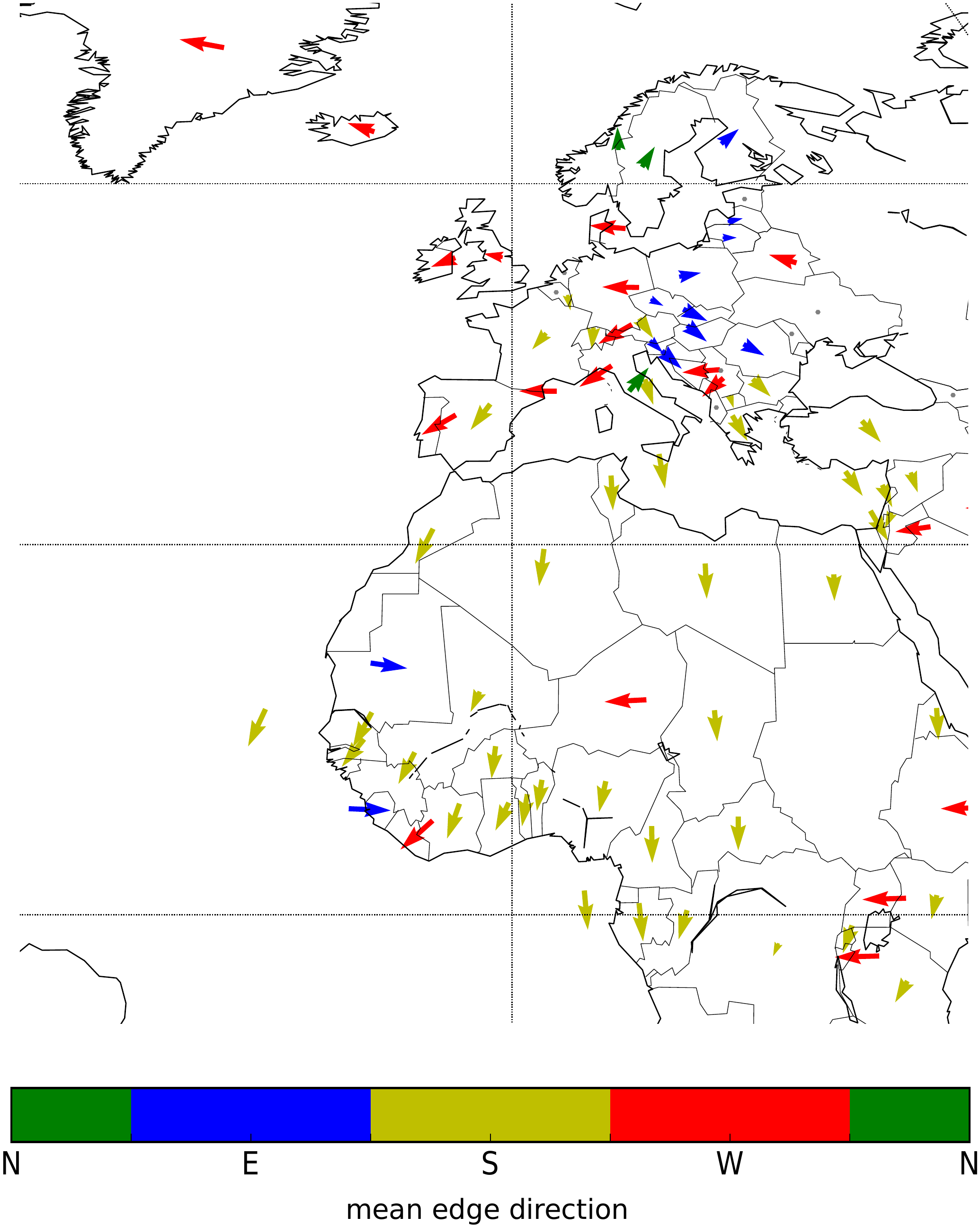}
\caption{Mean edge direction of an excerpt of the global import network including Europe and northern Africa.}\label{import_fig} 
\end{figure}

As the European Union is accountable for approximately one-third of the worldwide export, its role becomes especially visible in the import network (Fig.~\ref{import_fig}). The mean edge directions of countries from the Middle East, North and Central Africa and South America point mostly away from Europe as they primarily import goods from the EU member states. Another interesting feature is that most western EU countries (except for Austria and Italy) exhibit mean edge directions pointing westward and most eastern (post-2004) EU member states such pointing eastward. Of course, this is also induced by the large trade volume in the EU domestic market (all-to-all trade in a confined region automatically leads to converging/diverging edge directions), but it is very interesting, that the post eastward enlargement members are characterized by eastward (blue) pointing mean edge directions. In addition, one can distinguish non-EU members which are not well integrated into the EU domestic market, such as Serbia, Kosovo (west, red) and San Marino (north, green).

\section{Conclusions and Outlook}
We have studied edge directionality as a geometric network concept for analyzing spatially embedded networks with homogeneous and heterogeneous node placement. Specifically, we have utilized the mean edge direction and the local mean angle as geometric network characteristics and discussed their definitions in spherical geometry. Together with the previously introduced edge anisotropy \cite{Molkenthin2017} as well as metric properties associated with physical edge lengths, these measures complement spatial network analysis by a geometric perspective.

Motivated by the study of functional climate networks, we  have shown that angularly regular spherical grids induce a generic bias in both, local mean angle and edge anisotropy, which can be corrected by choosing edge weights according to the area of representation of the adjacent nodes. Even more, we argue that a similar bias necessarily appears for any type of heterogeneous areas of representation of nodes, e.g., when considering weighted networks based on general non-regular spherical graphs along with the corresponding Voronoi tessalation on the sphere. The bias correction approach proposed here is general enough to also cope with such more complex situations.

By applying our edge directionality measures to functional climate networks, we have demonstrated that the local edge anisotropy in surface air temperature (SAT) networks from idealized aquaplanet simulations reflects the structure of large-scale circulation cells. Mean edge direction and local edge anisotropy of networks generated from real-world ERA-Interim reanalysis SAT and precipitation data have demonstrated that nodes located at continental west coasts often exhibit  highly anisotropic edge directions in the SAT network, and have revealed the El Ni\~{n}o Southern Oscillation and the Atlantic Ni\~{n}o as significant structures with opposing mean edge directions in the precipitation network.
  
Last but not least, we have investigated the edge directionality in a global air transportation network and the world trade network from 2009. Our results underline the complexity of the global air transportation system, as we have identified various regions with distinct edge directionality. While the mean edge directions of outgoing flights in Europe point southward, we have observed  convergent edge directions in East Asia. In the import and export relations of the trade network, we have found two striking features: firstly, the difference between import and export relations, which are clearly distinguishable in a geometric network perspective, and secondly, the influence of the export of the European Union on global trade patterns, which is characterized by divergent mean edge directions. 

Our study has demonstrated that edge directionality measures complement classical network analysis, where often only topological characteristics are utilized. While this paper has mainly served to introduce the methodological framework and illustrate its applications to diverse examples of spatial networks, we have not attempted here to analyze the presented data sets comprehensively. Among others, an analysis of the climate data separately for each season or, more generally, a temporal view on dynamical changes of edge directionality patterns could provide meaningful insights into the specific networks studied. To study smaller components of spherical networks in greater detail and reveal interpretable information on local changes of edge directionality, we would further need to account for boundary effects resulting from the exclusion of information from outside the spatial domain of interest, which may strongly affect the latter \cite{Rheinwalt2012}. In the present work, we have not discussed these effects as we have investigated global networks only. Finally, we emphasize that geometric network measures can be defined for networks embedded in general (non-spherical and non-Euclidean) geometries, which remains an open research question so far. To this end, we outline corresponding further investigations and developments as subjects of future research.

\appendix
\section*{Appendix: Analytical expression for spherical course angles}

\begin{figure}[t]
\centering
\includegraphics[width=0.35\textwidth]{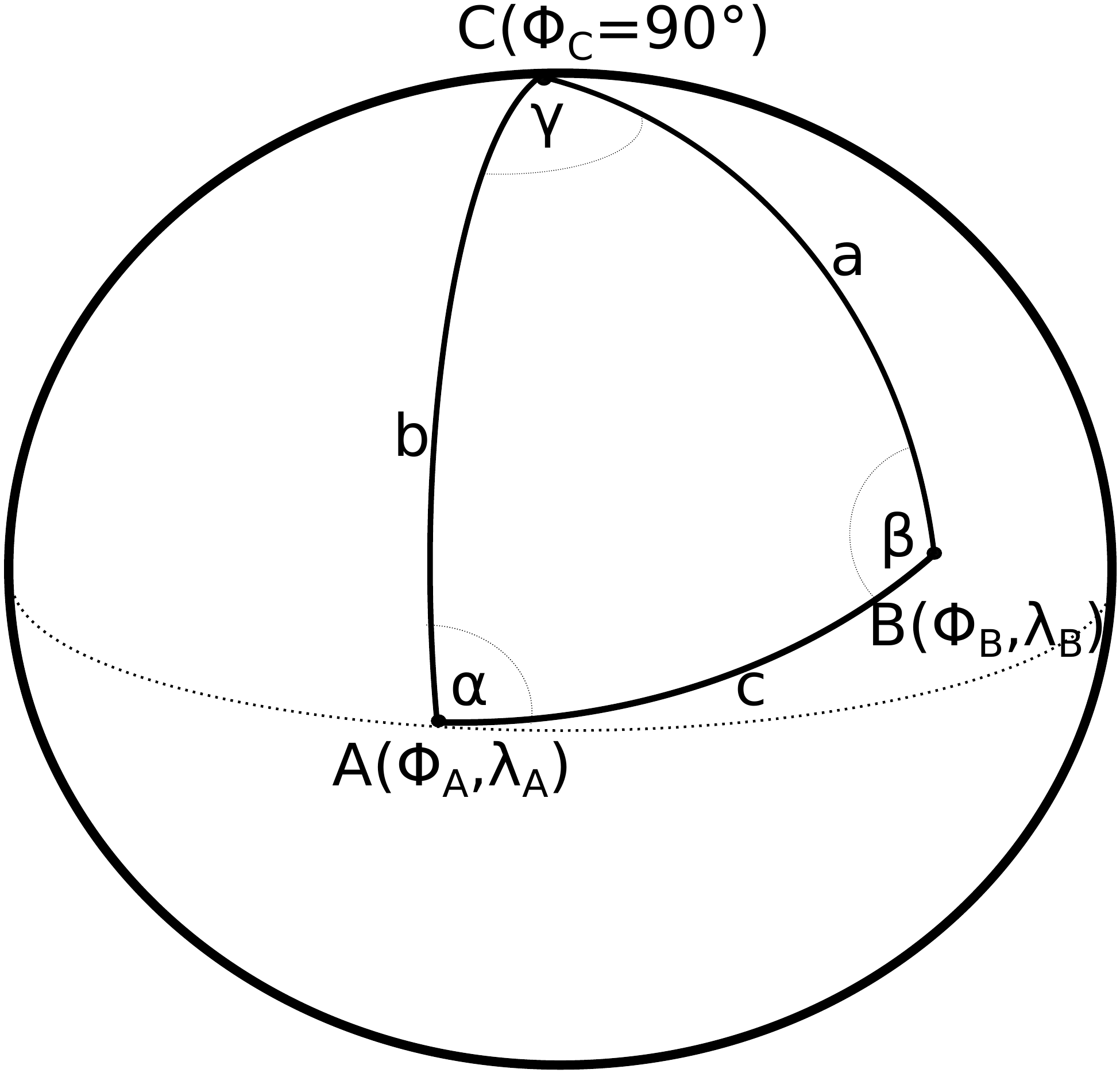}
\caption{Schematic illustration of a spherical triangle $\Delta ABC$ as utilized for the derivation of the course angle. Note that point $C$ is identified with the North Pole ($\phi_C=90^\circ$).}\label{triangle} 
\end{figure}

The derivation of analytical expressions for course angles requires utilizing basic concepts from spherical geometry. Let us consider an arbitrary spherical triangle with three sides of length $a$, $b$ and $c$, respectively. The three corners are labeled with $A$, $B$ and $C$ (such that the side with length $a$ is opposite to $A$, etc.), and are associated with angles $\alpha$, $\beta$ and $\gamma$, respectively. Put differently, $a$ is the geodetic distance between $B$ and $C$, etc. Figure~\ref{triangle} provides a corresponding schematic illustration. For convenience, the angular distances $a$, $b$ and $c$ are measured in radians on a unit sphere, and actual (physical) distances can be obtained by multiplying them with the radius of the considered sphere. 

First, we recall the spherical law of cosine (cosine rule for sides on a sphere) as
\begin{eqnarray}
\cos a &=& \cos b\cos c+\sin b \sin c \cos \alpha \nonumber \\
&\Leftrightarrow& \cos \alpha = \frac{\cos a -\cos b\cos c}{\sin b \sin c }. \label{eq:cosinerule}
\end{eqnarray}
\noindent 
For computing the angle between the side with length $c$ and the true northern direction we identify the point $C$ with the North Pole. 
We can now write the angular distance between $A$ and $C$ (and $B$ and $C$, respectively)  in terms of the latitudes $\phi_A$ and $\phi_B$ of $A$ and $B$ as $\cos a = \sin \phi_B$ and $ \sin b = \cos \phi_A$. Thus, we find
\begin{equation}
\cos \alpha = \frac{\sin \phi_B  -\sin \phi_A \cos c}{\cos \phi_A \sin c }
= \frac{\sin \phi_B  -\sin \phi_A \cos c}{\cos \phi_A \sqrt{1-\cos^2 c} }
. \label{eq:cosinealpha}
\end{equation}
\noindent
Note that the second identity in Eq.~\eqref{eq:cosinealpha} ignores the second possible (negative) solution of $\sin c=\pm \sqrt{1-\cos^2 c}$, which would lead to the supplement angle $\pi-\alpha$ that is not of interest here given our definition of course angles. 

In a similar fashion, the angular distance between the two points $A$ and $B$ can be expressed as
\begin{equation}
\cos c =\sin \phi_A \sin \phi_B + \cos \phi_A \cos \phi_B \cos (\lambda_B-\lambda_A), \label{eq:cosinedistance}
\end{equation}
which can be calculated in a straightforward manner from the scalar product in spherical coordinates between two vectors pointing towards $A$ and $B$.

Inserting Eq.~\eqref{eq:cosinedistance} into Eq.~\eqref{eq:cosinealpha} and making use of standard trigonometric identities results in the expression given in Eq.~\eqref{eq:courseangle} when identifying $\alpha$ with the course angle associated with the geodetic between $A$ and $B$.

\begin{acknowledgements}
This work has been financially supported by the IRTG 1740/TRP 2011/50151-0 (funded by the DFG and FAPESP) and by the German Federal Ministry for Education and Research (BMBF) via the BMBF Young Investigators Group $\text{CoSy-CC}^2$: Complex Systems Approaches to Understanding Causes and Consequences of Past, Present and Future Climate Change (grant no. 01LN1306A) and the Belmont Forum/JPI Climate project GOTHAM (grant no. 01LP16MA).

We thank Aiko Voigt for his guidance on the aquaplanet simulation data, Sven Willner and the Zeean research group for providing the world trade data, and two anonymous reviewers for their helpful comments on this manuscript. 

This version of this article has been resubmitted to \url{https://journals.aps.org/pre/} and has been uploaded within the APS Coyright Agreement Guidelines.
\end{acknowledgements}

\bibliography{library}

\begin{thebibliography}{41}%
\makeatletter
\providecommand \@ifxundefined [1]{%
 \@ifx{#1\undefined}
}%
\providecommand \@ifnum [1]{%
 \ifnum #1\expandafter \@firstoftwo
 \else \expandafter \@secondoftwo
 \fi
}%
\providecommand \@ifx [1]{%
 \ifx #1\expandafter \@firstoftwo
 \else \expandafter \@secondoftwo
 \fi
}%
\providecommand \natexlab [1]{#1}%
\providecommand \enquote  [1]{``#1''}%
\providecommand \bibnamefont  [1]{#1}%
\providecommand \bibfnamefont [1]{#1}%
\providecommand \citenamefont [1]{#1}%
\providecommand \href@noop [0]{\@secondoftwo}%
\providecommand \href [0]{\begingroup \@sanitize@url \@href}%
\providecommand \@href[1]{\@@startlink{#1}\@@href}%
\providecommand \@@href[1]{\endgroup#1\@@endlink}%
\providecommand \@sanitize@url [0]{\catcode `\\12\catcode `\$12\catcode
  `\&12\catcode `\#12\catcode `\^12\catcode `\_12\catcode `\%12\relax}%
\providecommand \@@startlink[1]{}%
\providecommand \@@endlink[0]{}%
\providecommand \url  [0]{\begingroup\@sanitize@url \@url }%
\providecommand \@url [1]{\endgroup\@href {#1}{\urlprefix }}%
\providecommand \urlprefix  [0]{URL }%
\providecommand \Eprint [0]{\href }%
\providecommand \doibase [0]{http://dx.doi.org/}%
\providecommand \selectlanguage [0]{\@gobble}%
\providecommand \bibinfo  [0]{\@secondoftwo}%
\providecommand \bibfield  [0]{\@secondoftwo}%
\providecommand \translation [1]{[#1]}%
\providecommand \BibitemOpen [0]{}%
\providecommand \bibitemStop [0]{}%
\providecommand \bibitemNoStop [0]{.\EOS\space}%
\providecommand \EOS [0]{\spacefactor3000\relax}%
\providecommand \BibitemShut  [1]{\csname bibitem#1\endcsname}%
\let\auto@bib@innerbib\@empty
\bibitem [{\citenamefont {Strogatz}(2001)}]{Strogatz2001}%
  \BibitemOpen
  \bibfield  {author} {\bibinfo {author} {\bibfnamefont {S.~H.}\ \bibnamefont
  {Strogatz}},\ }\href@noop {} {\bibfield  {journal} {\bibinfo  {journal}
  {Nature}\ }\textbf {\bibinfo {volume} {410}},\ \bibinfo {pages} {268}
  (\bibinfo {year} {2001})}\BibitemShut {NoStop}%
\bibitem [{\citenamefont {Albert}\ and\ \citenamefont
  {Barab{\'{a}}si}(2002)}]{Albert2002}%
  \BibitemOpen
  \bibfield  {author} {\bibinfo {author} {\bibfnamefont {R.}~\bibnamefont
  {Albert}}\ and\ \bibinfo {author} {\bibfnamefont {A.-L.}\ \bibnamefont
  {Barab{\'{a}}si}},\ }\href@noop {} {\bibfield  {journal} {\bibinfo  {journal}
  {Rev. Mod. Phys.}\ }\textbf {\bibinfo {volume} {74}},\ \bibinfo {pages} {47}
  (\bibinfo {year} {2002})}\BibitemShut {NoStop}%
\bibitem [{\citenamefont {Newman}(2003)}]{Newman2003}%
  \BibitemOpen
  \bibfield  {author} {\bibinfo {author} {\bibfnamefont {M.~E.~J.}\
  \bibnamefont {Newman}},\ }\href@noop {} {\bibfield  {journal} {\bibinfo
  {journal} {SIAM Rev.}\ }\textbf {\bibinfo {volume} {45}},\ \bibinfo {pages}
  {167} (\bibinfo {year} {2003})}\BibitemShut {NoStop}%
\bibitem [{\citenamefont {Boccaletti}\ \emph {et~al.}(2006)\citenamefont
  {Boccaletti}, \citenamefont {Latora}, \citenamefont {Moreno}, \citenamefont
  {Chavez},\ and\ \citenamefont {Hwang}}]{Boccaletti2006}%
  \BibitemOpen
  \bibfield  {author} {\bibinfo {author} {\bibfnamefont {S.}~\bibnamefont
  {Boccaletti}}, \bibinfo {author} {\bibfnamefont {V.}~\bibnamefont {Latora}},
  \bibinfo {author} {\bibfnamefont {Y.}~\bibnamefont {Moreno}}, \bibinfo
  {author} {\bibfnamefont {M.}~\bibnamefont {Chavez}}, \ and\ \bibinfo {author}
  {\bibfnamefont {D.~U.}\ \bibnamefont {Hwang}},\ }\href@noop {} {\bibfield
  {journal} {\bibinfo  {journal} {Phys. Rep.}\ }\textbf {\bibinfo {volume}
  {424}},\ \bibinfo {pages} {175} (\bibinfo {year} {2006})}\BibitemShut
  {NoStop}%
\bibitem [{\citenamefont {Dunne}\ \emph {et~al.}(2002)\citenamefont {Dunne},
  \citenamefont {Williams},\ and\ \citenamefont {Martinez}}]{Dunne2002}%
  \BibitemOpen
  \bibfield  {author} {\bibinfo {author} {\bibfnamefont {J.~A.}\ \bibnamefont
  {Dunne}}, \bibinfo {author} {\bibfnamefont {R.~J.}\ \bibnamefont {Williams}},
  \ and\ \bibinfo {author} {\bibfnamefont {N.~D.}\ \bibnamefont {Martinez}},\
  }\href@noop {} {\bibfield  {journal} {\bibinfo  {journal} {Ecol. Lett.}\
  }\textbf {\bibinfo {volume} {5}},\ \bibinfo {pages} {558} (\bibinfo {year}
  {2002})}\BibitemShut {NoStop}%
\bibitem [{\citenamefont {Girvan}\ and\ \citenamefont
  {Newman}(2002)}]{Girvan2002}%
  \BibitemOpen
  \bibfield  {author} {\bibinfo {author} {\bibfnamefont {M.}~\bibnamefont
  {Girvan}}\ and\ \bibinfo {author} {\bibfnamefont {M.~E.~J.}\ \bibnamefont
  {Newman}},\ }\href@noop {} {\bibfield  {journal} {\bibinfo  {journal} {Proc.
  Natl. Acad. Sci.}\ }\textbf {\bibinfo {volume} {99}},\ \bibinfo {pages}
  {7821} (\bibinfo {year} {2002})}\BibitemShut {NoStop}%
\bibitem [{\citenamefont {Tsonis}\ and\ \citenamefont
  {Roebber}(2004)}]{Tsonis2004}%
  \BibitemOpen
  \bibfield  {author} {\bibinfo {author} {\bibfnamefont {A.~A.}\ \bibnamefont
  {Tsonis}}\ and\ \bibinfo {author} {\bibfnamefont {P.~J.}\ \bibnamefont
  {Roebber}},\ }\href@noop {} {\bibfield  {journal} {\bibinfo  {journal} {Phys.
  A}\ }\textbf {\bibinfo {volume} {333}},\ \bibinfo {pages} {497} (\bibinfo
  {year} {2004})}\BibitemShut {NoStop}%
\bibitem [{\citenamefont {Zhou}\ \emph {et~al.}(2006)\citenamefont {Zhou},
  \citenamefont {Zemanov{\'{a}}}, \citenamefont {Zamora}, \citenamefont
  {Hilgetag},\ and\ \citenamefont {Kurths}}]{Zhou2006}%
  \BibitemOpen
  \bibfield  {author} {\bibinfo {author} {\bibfnamefont {C.}~\bibnamefont
  {Zhou}}, \bibinfo {author} {\bibfnamefont {L.}~\bibnamefont
  {Zemanov{\'{a}}}}, \bibinfo {author} {\bibfnamefont {G.}~\bibnamefont
  {Zamora}}, \bibinfo {author} {\bibfnamefont {C.~C.}\ \bibnamefont
  {Hilgetag}}, \ and\ \bibinfo {author} {\bibfnamefont {J.}~\bibnamefont
  {Kurths}},\ }\href@noop {} {\bibfield  {journal} {\bibinfo  {journal} {Phys.
  Rev. Lett.}\ }\textbf {\bibinfo {volume} {97}},\ \bibinfo {pages} {238103}
  (\bibinfo {year} {2006})}\BibitemShut {NoStop}%
\bibitem [{\citenamefont {Landherr}\ \emph {et~al.}(2010)\citenamefont
  {Landherr}, \citenamefont {Friedl},\ and\ \citenamefont
  {Heidemann}}]{Landherr2010}%
  \BibitemOpen
  \bibfield  {author} {\bibinfo {author} {\bibfnamefont {A.}~\bibnamefont
  {Landherr}}, \bibinfo {author} {\bibfnamefont {B.}~\bibnamefont {Friedl}}, \
  and\ \bibinfo {author} {\bibfnamefont {J.}~\bibnamefont {Heidemann}},\
  }\href@noop {} {\bibfield  {journal} {\bibinfo  {journal} {Bus. Inf. Syst.
  Eng.}\ }\textbf {\bibinfo {volume} {2}},\ \bibinfo {pages} {371} (\bibinfo
  {year} {2010})}\BibitemShut {NoStop}%
\bibitem [{\citenamefont {Gastner}\ and\ \citenamefont
  {Newman}(2006)}]{Gastner2006}%
  \BibitemOpen
  \bibfield  {author} {\bibinfo {author} {\bibfnamefont {M.~T.}\ \bibnamefont
  {Gastner}}\ and\ \bibinfo {author} {\bibfnamefont {M.~E.~J.}\ \bibnamefont
  {Newman}},\ }\href@noop {} {\bibfield  {journal} {\bibinfo  {journal} {Eur.
  Phys. J. B}\ }\textbf {\bibinfo {volume} {49}},\ \bibinfo {pages} {247}
  (\bibinfo {year} {2006})}\BibitemShut {NoStop}%
\bibitem [{\citenamefont {Masucci}\ \emph {et~al.}(2009)\citenamefont
  {Masucci}, \citenamefont {Smith}, \citenamefont {Crooks},\ and\ \citenamefont
  {Batty}}]{Masucci2009}%
  \BibitemOpen
  \bibfield  {author} {\bibinfo {author} {\bibfnamefont {A.~P.}\ \bibnamefont
  {Masucci}}, \bibinfo {author} {\bibfnamefont {D.}~\bibnamefont {Smith}},
  \bibinfo {author} {\bibfnamefont {A.}~\bibnamefont {Crooks}}, \ and\ \bibinfo
  {author} {\bibfnamefont {M.}~\bibnamefont {Batty}},\ }\href@noop {}
  {\bibfield  {journal} {\bibinfo  {journal} {Eur. Phys. J. B}\ }\textbf
  {\bibinfo {volume} {71}},\ \bibinfo {pages} {259} (\bibinfo {year}
  {2009})}\BibitemShut {NoStop}%
\bibitem [{\citenamefont {Barthelemy}(2011)}]{Barthelemy2011}%
  \BibitemOpen
  \bibfield  {author} {\bibinfo {author} {\bibfnamefont {M.}~\bibnamefont
  {Barthelemy}},\ }\href@noop {} {\bibfield  {journal} {\bibinfo  {journal}
  {Phys. Rep.}\ }\textbf {\bibinfo {volume} {499}},\ \bibinfo {pages} {1}
  (\bibinfo {year} {2011})}\BibitemShut {NoStop}%
\bibitem [{\citenamefont {Bialonski}\ \emph {et~al.}(2010)\citenamefont
  {Bialonski}, \citenamefont {Horstmann},\ and\ \citenamefont
  {Lehnertz}}]{Bialonski2010}%
  \BibitemOpen
  \bibfield  {author} {\bibinfo {author} {\bibfnamefont {S.}~\bibnamefont
  {Bialonski}}, \bibinfo {author} {\bibfnamefont {M.~T.}\ \bibnamefont
  {Horstmann}}, \ and\ \bibinfo {author} {\bibfnamefont {K.}~\bibnamefont
  {Lehnertz}},\ }\href@noop {} {\bibfield  {journal} {\bibinfo  {journal}
  {Chaos}\ }\textbf {\bibinfo {volume} {20}},\ \bibinfo {pages} {013134}
  (\bibinfo {year} {2010})}\BibitemShut {NoStop}%
\bibitem [{\citenamefont {Chan}\ \emph {et~al.}(2011)\citenamefont {Chan},
  \citenamefont {Donner},\ and\ \citenamefont {Laemmer}}]{Chan2011}%
  \BibitemOpen
  \bibfield  {author} {\bibinfo {author} {\bibfnamefont {S.~H.~Y.}\
  \bibnamefont {Chan}}, \bibinfo {author} {\bibfnamefont {R.~V.}\ \bibnamefont
  {Donner}}, \ and\ \bibinfo {author} {\bibfnamefont {S.}~\bibnamefont
  {Laemmer}},\ }\href@noop {} {\bibfield  {journal} {\bibinfo  {journal} {Eur.
  Phys. J. B}\ }\textbf {\bibinfo {volume} {84}},\ \bibinfo {pages} {563}
  (\bibinfo {year} {2011})}\BibitemShut {NoStop}%
\bibitem [{\citenamefont {Schultz}\ \emph {et~al.}(2014)\citenamefont
  {Schultz}, \citenamefont {Heitzig},\ and\ \citenamefont
  {Kurths}}]{Schultz2014}%
  \BibitemOpen
  \bibfield  {author} {\bibinfo {author} {\bibfnamefont {P.}~\bibnamefont
  {Schultz}}, \bibinfo {author} {\bibfnamefont {J.}~\bibnamefont {Heitzig}}, \
  and\ \bibinfo {author} {\bibfnamefont {J.}~\bibnamefont {Kurths}},\
  }\href@noop {} {\bibfield  {journal} {\bibinfo  {journal} {EPL}\ }\textbf
  {\bibinfo {volume} {223}},\ \bibinfo {pages} {2593} (\bibinfo {year}
  {2014})}\BibitemShut {NoStop}%
\bibitem [{\citenamefont {Wiedermann}\ \emph
  {et~al.}(2017{\natexlab{a}})\citenamefont {Wiedermann}, \citenamefont
  {Donges}, \citenamefont {Kurths},\ and\ \citenamefont
  {Donner}}]{Wiedermann2017c}%
  \BibitemOpen
  \bibfield  {author} {\bibinfo {author} {\bibfnamefont {M.}~\bibnamefont
  {Wiedermann}}, \bibinfo {author} {\bibfnamefont {J.~F.}\ \bibnamefont
  {Donges}}, \bibinfo {author} {\bibfnamefont {J.}~\bibnamefont {Kurths}}, \
  and\ \bibinfo {author} {\bibfnamefont {R.~V.}\ \bibnamefont {Donner}},\
  }\href@noop {} {\bibfield  {journal} {\bibinfo  {journal} {Phys. Rev. E}\
  }\textbf {\bibinfo {volume} {96}},\ \bibinfo {pages} {042304} (\bibinfo
  {year} {2017}{\natexlab{a}})}\BibitemShut {NoStop}%
\bibitem [{\citenamefont {Molkenthin}\ \emph {et~al.}(2017)\citenamefont
  {Molkenthin}, \citenamefont {Kutza}, \citenamefont {Tupikina}, \citenamefont
  {Marwan}, \citenamefont {Donges}, \citenamefont {Feudel},\ and\ \citenamefont
  {Donner}}]{Molkenthin2017}%
  \BibitemOpen
  \bibfield  {author} {\bibinfo {author} {\bibfnamefont {N.}~\bibnamefont
  {Molkenthin}}, \bibinfo {author} {\bibfnamefont {H.}~\bibnamefont {Kutza}},
  \bibinfo {author} {\bibfnamefont {L.}~\bibnamefont {Tupikina}}, \bibinfo
  {author} {\bibfnamefont {N.}~\bibnamefont {Marwan}}, \bibinfo {author}
  {\bibfnamefont {J.~F.}\ \bibnamefont {Donges}}, \bibinfo {author}
  {\bibfnamefont {U.}~\bibnamefont {Feudel}}, \ and\ \bibinfo {author}
  {\bibfnamefont {R.~V.}\ \bibnamefont {Donner}},\ }\href@noop {} {\bibfield
  {journal} {\bibinfo  {journal} {Chaos}\ }\textbf {\bibinfo {volume} {27}},\
  \bibinfo {pages} {035802} (\bibinfo {year} {2017})}\BibitemShut {NoStop}%
\bibitem [{\citenamefont {Banavar}\ \emph {et~al.}(1999)\citenamefont
  {Banavar}, \citenamefont {Maritan},\ and\ \citenamefont
  {Rinaldo}}]{Banavar1999}%
  \BibitemOpen
  \bibfield  {author} {\bibinfo {author} {\bibfnamefont {J.~R.}\ \bibnamefont
  {Banavar}}, \bibinfo {author} {\bibfnamefont {A.}~\bibnamefont {Maritan}}, \
  and\ \bibinfo {author} {\bibfnamefont {A.}~\bibnamefont {Rinaldo}},\
  }\href@noop {} {\bibfield  {journal} {\bibinfo  {journal} {Nature}\ }\textbf
  {\bibinfo {volume} {399}},\ \bibinfo {pages} {130} (\bibinfo {year}
  {1999})}\BibitemShut {NoStop}%
\bibitem [{\citenamefont {Serrano}\ and\ \citenamefont
  {Bogu{\~{n}}{\'{a}}}(2003)}]{Serrano2003}%
  \BibitemOpen
  \bibfield  {author} {\bibinfo {author} {\bibfnamefont {M.~{\'{A}}.}\
  \bibnamefont {Serrano}}\ and\ \bibinfo {author} {\bibfnamefont
  {M.}~\bibnamefont {Bogu{\~{n}}{\'{a}}}},\ }\href@noop {} {\bibfield
  {journal} {\bibinfo  {journal} {Phys. Rev. E}\ }\textbf {\bibinfo {volume}
  {68}},\ \bibinfo {pages} {015101} (\bibinfo {year} {2003})}\BibitemShut
  {NoStop}%
\bibitem [{\citenamefont {Achard}(2006)}]{Achard2006}%
  \BibitemOpen
  \bibfield  {author} {\bibinfo {author} {\bibfnamefont {S.}~\bibnamefont
  {Achard}},\ }\href@noop {} {\bibfield  {journal} {\bibinfo  {journal} {J.
  Neurosci.}\ }\textbf {\bibinfo {volume} {26}},\ \bibinfo {pages} {63}
  (\bibinfo {year} {2006})}\BibitemShut {NoStop}%
\bibitem [{\citenamefont {Donges}\ \emph {et~al.}(2009)\citenamefont {Donges},
  \citenamefont {Zou}, \citenamefont {Marwan},\ and\ \citenamefont
  {Kurths}}]{Donges2009}%
  \BibitemOpen
  \bibfield  {author} {\bibinfo {author} {\bibfnamefont {J.~F.}\ \bibnamefont
  {Donges}}, \bibinfo {author} {\bibfnamefont {Y.}~\bibnamefont {Zou}},
  \bibinfo {author} {\bibfnamefont {N.}~\bibnamefont {Marwan}}, \ and\ \bibinfo
  {author} {\bibfnamefont {J.}~\bibnamefont {Kurths}},\ }\href@noop {}
  {\bibfield  {journal} {\bibinfo  {journal} {EPL}\ }\textbf {\bibinfo {volume}
  {87}},\ \bibinfo {pages} {48007} (\bibinfo {year} {2009})}\BibitemShut
  {NoStop}%
\bibitem [{\citenamefont {Donner}\ \emph {et~al.}(2017)\citenamefont {Donner},
  \citenamefont {Wiedermann},\ and\ \citenamefont {Donges}}]{Donner2017}%
  \BibitemOpen
  \bibfield  {author} {\bibinfo {author} {\bibfnamefont {R.~V.}\ \bibnamefont
  {Donner}}, \bibinfo {author} {\bibfnamefont {M.}~\bibnamefont {Wiedermann}},
  \ and\ \bibinfo {author} {\bibfnamefont {J.~F.}\ \bibnamefont {Donges}},\
  }\href@noop {} {\emph {\bibinfo {title} {Nonlinear Stoch. Clim. Dyn.}}},\
  \bibinfo {edition} {1st}\ ed.,\ edited by\ \bibinfo {editor} {\bibfnamefont
  {C.}~\bibnamefont {Franzke}}\ and\ \bibinfo {editor} {\bibfnamefont
  {T.}~\bibnamefont {O'Kane}}\ (\bibinfo  {publisher} {Cambridge University
  Press},\ \bibinfo {address} {Cambridge},\ \bibinfo {year} {2017})\ pp.\
  \bibinfo {pages} {159--183}\BibitemShut {NoStop}%
\bibitem [{\citenamefont {Heitzig}\ \emph {et~al.}(2012)\citenamefont
  {Heitzig}, \citenamefont {Donges}, \citenamefont {Zou}, \citenamefont
  {Marwan},\ and\ \citenamefont {Kurths}}]{Heitzig2012}%
  \BibitemOpen
  \bibfield  {author} {\bibinfo {author} {\bibfnamefont {J.}~\bibnamefont
  {Heitzig}}, \bibinfo {author} {\bibfnamefont {J.~F.}\ \bibnamefont {Donges}},
  \bibinfo {author} {\bibfnamefont {Y.}~\bibnamefont {Zou}}, \bibinfo {author}
  {\bibfnamefont {N.}~\bibnamefont {Marwan}}, \ and\ \bibinfo {author}
  {\bibfnamefont {J.}~\bibnamefont {Kurths}},\ }\href@noop {} {\bibfield
  {journal} {\bibinfo  {journal} {Eur. Phys. J. B}\ }\textbf {\bibinfo {volume}
  {85}},\ \bibinfo {pages} {38} (\bibinfo {year} {2012})}\BibitemShut {NoStop}%
\bibitem [{\citenamefont {Wiedermann}\ \emph {et~al.}(2013)\citenamefont
  {Wiedermann}, \citenamefont {Donges}, \citenamefont {Heitzig},\ and\
  \citenamefont {Kurths}}]{Wiedermann2013}%
  \BibitemOpen
  \bibfield  {author} {\bibinfo {author} {\bibfnamefont {M.}~\bibnamefont
  {Wiedermann}}, \bibinfo {author} {\bibfnamefont {J.~F.}\ \bibnamefont
  {Donges}}, \bibinfo {author} {\bibfnamefont {J.}~\bibnamefont {Heitzig}}, \
  and\ \bibinfo {author} {\bibfnamefont {J.}~\bibnamefont {Kurths}},\
  }\href@noop {} {\bibfield  {journal} {\bibinfo  {journal} {EPL}\ }\textbf
  {\bibinfo {volume} {102}},\ \bibinfo {pages} {28007} (\bibinfo {year}
  {2013})}\BibitemShut {NoStop}%
\bibitem [{\citenamefont {Zemp}\ \emph {et~al.}(2014)\citenamefont {Zemp},
  \citenamefont {Wiedermann}, \citenamefont {Kurths}, \citenamefont {Rammig},\
  and\ \citenamefont {Donges}}]{Zemp2014}%
  \BibitemOpen
  \bibfield  {author} {\bibinfo {author} {\bibfnamefont {D.~C.}\ \bibnamefont
  {Zemp}}, \bibinfo {author} {\bibfnamefont {M.}~\bibnamefont {Wiedermann}},
  \bibinfo {author} {\bibfnamefont {J.}~\bibnamefont {Kurths}}, \bibinfo
  {author} {\bibfnamefont {A.}~\bibnamefont {Rammig}}, \ and\ \bibinfo {author}
  {\bibfnamefont {J.~F.}\ \bibnamefont {Donges}},\ }\href@noop {} {\bibfield
  {journal} {\bibinfo  {journal} {EPL}\ }\textbf {\bibinfo {volume} {107}},\
  \bibinfo {pages} {58005} (\bibinfo {year} {2014})}\BibitemShut {NoStop}%
\bibitem [{\citenamefont {Wiedermann}\ \emph
  {et~al.}(2017{\natexlab{b}})\citenamefont {Wiedermann}, \citenamefont
  {Donges}, \citenamefont {Handorf}, \citenamefont {Kurths},\ and\
  \citenamefont {Donner}}]{Wiedermann2017}%
  \BibitemOpen
  \bibfield  {author} {\bibinfo {author} {\bibfnamefont {M.}~\bibnamefont
  {Wiedermann}}, \bibinfo {author} {\bibfnamefont {J.~F.}\ \bibnamefont
  {Donges}}, \bibinfo {author} {\bibfnamefont {D.}~\bibnamefont {Handorf}},
  \bibinfo {author} {\bibfnamefont {J.}~\bibnamefont {Kurths}}, \ and\ \bibinfo
  {author} {\bibfnamefont {R.~V.}\ \bibnamefont {Donner}},\ }\href@noop {}
  {\bibfield  {journal} {\bibinfo  {journal} {Int. J. Climatol.}\ }\textbf
  {\bibinfo {volume} {37}},\ \bibinfo {pages} {3821} (\bibinfo {year}
  {2017}{\natexlab{b}})}\BibitemShut {NoStop}%
\bibitem [{\citenamefont {Gudmundsson}\ and\ \citenamefont
  {Mohajeri}(2013)}]{Gudmundsson2013}%
  \BibitemOpen
  \bibfield  {author} {\bibinfo {author} {\bibfnamefont {A.}~\bibnamefont
  {Gudmundsson}}\ and\ \bibinfo {author} {\bibfnamefont {N.}~\bibnamefont
  {Mohajeri}},\ }\href@noop {} {\bibfield  {journal} {\bibinfo  {journal} {Sci.
  Rep.}\ }\textbf {\bibinfo {volume} {3}},\ \bibinfo {pages} {3324} (\bibinfo
  {year} {2013})}\BibitemShut {NoStop}%
\bibitem [{\citenamefont {Mohajeri}\ \emph {et~al.}(2013)\citenamefont
  {Mohajeri}, \citenamefont {French},\ and\ \citenamefont
  {Gudmundsson}}]{Mohajeri2013}%
  \BibitemOpen
  \bibfield  {author} {\bibinfo {author} {\bibfnamefont {N.}~\bibnamefont
  {Mohajeri}}, \bibinfo {author} {\bibfnamefont {J.~R.}\ \bibnamefont
  {French}}, \ and\ \bibinfo {author} {\bibfnamefont {A.}~\bibnamefont
  {Gudmundsson}},\ }\href@noop {} {\bibfield  {journal} {\bibinfo  {journal}
  {Entropy}\ }\textbf {\bibinfo {volume} {15}},\ \bibinfo {pages} {3340}
  (\bibinfo {year} {2013})}\BibitemShut {NoStop}%
\bibitem [{\citenamefont {Boers}\ and\ \citenamefont
  {Rheinwalt}(2014)}]{Boers2014a}%
  \BibitemOpen
  \bibfield  {author} {\bibinfo {author} {\bibfnamefont {N.}~\bibnamefont
  {Boers}}\ and\ \bibinfo {author} {\bibfnamefont {A.}~\bibnamefont
  {Rheinwalt}},\ }\href@noop {} {\bibfield  {journal} {\bibinfo  {journal}
  {Geophys. Res. Lett.}\ }\textbf {\bibinfo {volume} {41}},\ \bibinfo {pages}
  {7397} (\bibinfo {year} {2014})}\BibitemShut {NoStop}%
\bibitem [{\citenamefont {Rheinwalt}\ \emph {et~al.}(2016)\citenamefont
  {Rheinwalt}, \citenamefont {Boers}, \citenamefont {Marwan}, \citenamefont
  {Kurths}, \citenamefont {Hoffmann}, \citenamefont {Gerstengarbe},\ and\
  \citenamefont {Werner}}]{Rheinwalt2016}%
  \BibitemOpen
  \bibfield  {author} {\bibinfo {author} {\bibfnamefont {A.}~\bibnamefont
  {Rheinwalt}}, \bibinfo {author} {\bibfnamefont {N.}~\bibnamefont {Boers}},
  \bibinfo {author} {\bibfnamefont {N.}~\bibnamefont {Marwan}}, \bibinfo
  {author} {\bibfnamefont {J.}~\bibnamefont {Kurths}}, \bibinfo {author}
  {\bibfnamefont {P.}~\bibnamefont {Hoffmann}}, \bibinfo {author}
  {\bibfnamefont {F.~W.}\ \bibnamefont {Gerstengarbe}}, \ and\ \bibinfo
  {author} {\bibfnamefont {P.}~\bibnamefont {Werner}},\ }\href@noop {}
  {\bibfield  {journal} {\bibinfo  {journal} {Clim. Dyn.}\ }\textbf {\bibinfo
  {volume} {46}},\ \bibinfo {pages} {1065} (\bibinfo {year}
  {2016})}\BibitemShut {NoStop}%
\bibitem [{\citenamefont {Radebach}\ \emph {et~al.}(2013)\citenamefont
  {Radebach}, \citenamefont {Donner}, \citenamefont {Runge}, \citenamefont
  {Donges},\ and\ \citenamefont {Kurths}}]{Radebach2013}%
  \BibitemOpen
  \bibfield  {author} {\bibinfo {author} {\bibfnamefont {A.}~\bibnamefont
  {Radebach}}, \bibinfo {author} {\bibfnamefont {R.~V.}\ \bibnamefont
  {Donner}}, \bibinfo {author} {\bibfnamefont {J.}~\bibnamefont {Runge}},
  \bibinfo {author} {\bibfnamefont {J.~F.}\ \bibnamefont {Donges}}, \ and\
  \bibinfo {author} {\bibfnamefont {J.}~\bibnamefont {Kurths}},\ }\href@noop {}
  {\bibfield  {journal} {\bibinfo  {journal} {Phys. Rev. E}\ }\textbf {\bibinfo
  {volume} {88}},\ \bibinfo {pages} {052807} (\bibinfo {year}
  {2013})}\BibitemShut {NoStop}%
\bibitem [{\citenamefont {Tsonis}\ \emph {et~al.}(2006)\citenamefont {Tsonis},
  \citenamefont {Swanson},\ and\ \citenamefont {Roebber}}]{Tsonis2006}%
  \BibitemOpen
  \bibfield  {author} {\bibinfo {author} {\bibfnamefont {A.~A.}\ \bibnamefont
  {Tsonis}}, \bibinfo {author} {\bibfnamefont {K.~L.}\ \bibnamefont {Swanson}},
  \ and\ \bibinfo {author} {\bibfnamefont {P.~J.}\ \bibnamefont {Roebber}},\
  }\href@noop {} {\bibfield  {journal} {\bibinfo  {journal} {Bull. Am.
  Meteorol. Soc.}\ }\textbf {\bibinfo {volume} {87}},\ \bibinfo {pages} {585}
  (\bibinfo {year} {2006})}\BibitemShut {NoStop}%
\bibitem [{\citenamefont {Voigt}\ \emph {et~al.}(2016)\citenamefont {Voigt},
  \citenamefont {Biasutti}, \citenamefont {Scheff}, \citenamefont {Bader},
  \citenamefont {Bordoni}, \citenamefont {Codron}, \citenamefont {Dixon},
  \citenamefont {Jonas}, \citenamefont {Kang}, \citenamefont {Klingaman},
  \citenamefont {Leung}, \citenamefont {Lu}, \citenamefont {Mapes},
  \citenamefont {Maroon}, \citenamefont {McDermid}, \citenamefont {Park},
  \citenamefont {Roehrig}, \citenamefont {Rose}, \citenamefont {Russell},
  \citenamefont {Seo}, \citenamefont {Toniazzo}, \citenamefont {Wei},
  \citenamefont {Yoshimori},\ and\ \citenamefont {{Vargas
  Zeppetello}}}]{Voigt2016}%
  \BibitemOpen
  \bibfield  {author} {\bibinfo {author} {\bibfnamefont {A.}~\bibnamefont
  {Voigt}}, \bibinfo {author} {\bibfnamefont {M.}~\bibnamefont {Biasutti}},
  \bibinfo {author} {\bibfnamefont {J.}~\bibnamefont {Scheff}}, \bibinfo
  {author} {\bibfnamefont {J.}~\bibnamefont {Bader}}, \bibinfo {author}
  {\bibfnamefont {S.}~\bibnamefont {Bordoni}}, \bibinfo {author} {\bibfnamefont
  {F.}~\bibnamefont {Codron}}, \bibinfo {author} {\bibfnamefont {R.~D.}\
  \bibnamefont {Dixon}}, \bibinfo {author} {\bibfnamefont {J.}~\bibnamefont
  {Jonas}}, \bibinfo {author} {\bibfnamefont {S.~M.}\ \bibnamefont {Kang}},
  \bibinfo {author} {\bibfnamefont {N.~P.}\ \bibnamefont {Klingaman}}, \bibinfo
  {author} {\bibfnamefont {R.}~\bibnamefont {Leung}}, \bibinfo {author}
  {\bibfnamefont {J.}~\bibnamefont {Lu}}, \bibinfo {author} {\bibfnamefont
  {B.}~\bibnamefont {Mapes}}, \bibinfo {author} {\bibfnamefont {E.~A.}\
  \bibnamefont {Maroon}}, \bibinfo {author} {\bibfnamefont {S.}~\bibnamefont
  {McDermid}}, \bibinfo {author} {\bibfnamefont {J.}~\bibnamefont {Park}},
  \bibinfo {author} {\bibfnamefont {R.}~\bibnamefont {Roehrig}}, \bibinfo
  {author} {\bibfnamefont {B.~E.~J.}\ \bibnamefont {Rose}}, \bibinfo {author}
  {\bibfnamefont {G.~L.}\ \bibnamefont {Russell}}, \bibinfo {author}
  {\bibfnamefont {J.}~\bibnamefont {Seo}}, \bibinfo {author} {\bibfnamefont
  {T.}~\bibnamefont {Toniazzo}}, \bibinfo {author} {\bibfnamefont
  {H.}~\bibnamefont {Wei}}, \bibinfo {author} {\bibfnamefont {M.}~\bibnamefont
  {Yoshimori}}, \ and\ \bibinfo {author} {\bibfnamefont {L.~R.}\ \bibnamefont
  {{Vargas Zeppetello}}},\ }\href@noop {} {\bibfield  {journal} {\bibinfo
  {journal} {J. Adv. Model. Earth Syst.}\ }\textbf {\bibinfo {volume} {8}},\
  \bibinfo {pages} {1868} (\bibinfo {year} {2016})}\BibitemShut {NoStop}%
\bibitem [{\citenamefont {Stevens}\ \emph {et~al.}(2013)\citenamefont
  {Stevens}, \citenamefont {Giorgetta}, \citenamefont {Esch}, \citenamefont
  {Mauritsen}, \citenamefont {Crueger}, \citenamefont {Rast}, \citenamefont
  {Salzmann}, \citenamefont {Schmidt}, \citenamefont {Bader}, \citenamefont
  {Block}, \citenamefont {Brokopf}, \citenamefont {Fast}, \citenamefont
  {Kinne}, \citenamefont {Kornblueh}, \citenamefont {Lohmann}, \citenamefont
  {Pincus}, \citenamefont {Reichler},\ and\ \citenamefont
  {Roeckner}}]{Stevens2013}%
  \BibitemOpen
  \bibfield  {author} {\bibinfo {author} {\bibfnamefont {B.}~\bibnamefont
  {Stevens}}, \bibinfo {author} {\bibfnamefont {M.}~\bibnamefont {Giorgetta}},
  \bibinfo {author} {\bibfnamefont {M.}~\bibnamefont {Esch}}, \bibinfo {author}
  {\bibfnamefont {T.}~\bibnamefont {Mauritsen}}, \bibinfo {author}
  {\bibfnamefont {T.}~\bibnamefont {Crueger}}, \bibinfo {author} {\bibfnamefont
  {S.}~\bibnamefont {Rast}}, \bibinfo {author} {\bibfnamefont {M.}~\bibnamefont
  {Salzmann}}, \bibinfo {author} {\bibfnamefont {H.}~\bibnamefont {Schmidt}},
  \bibinfo {author} {\bibfnamefont {J.}~\bibnamefont {Bader}}, \bibinfo
  {author} {\bibfnamefont {K.}~\bibnamefont {Block}}, \bibinfo {author}
  {\bibfnamefont {R.}~\bibnamefont {Brokopf}}, \bibinfo {author} {\bibfnamefont
  {I.}~\bibnamefont {Fast}}, \bibinfo {author} {\bibfnamefont {S.}~\bibnamefont
  {Kinne}}, \bibinfo {author} {\bibfnamefont {L.}~\bibnamefont {Kornblueh}},
  \bibinfo {author} {\bibfnamefont {U.}~\bibnamefont {Lohmann}}, \bibinfo
  {author} {\bibfnamefont {R.}~\bibnamefont {Pincus}}, \bibinfo {author}
  {\bibfnamefont {T.}~\bibnamefont {Reichler}}, \ and\ \bibinfo {author}
  {\bibfnamefont {E.}~\bibnamefont {Roeckner}},\ }\href@noop {} {\bibfield
  {journal} {\bibinfo  {journal} {J. Adv. Model. Earth Syst.}\ }\textbf
  {\bibinfo {volume} {5}},\ \bibinfo {pages} {146} (\bibinfo {year}
  {2013})}\BibitemShut {NoStop}%
\bibitem [{\citenamefont {Dee}\ \emph {et~al.}(2011)\citenamefont {Dee},
  \citenamefont {Uppala}, \citenamefont {Simmons}, \citenamefont {Berrisford},
  \citenamefont {Poli}, \citenamefont {Kobayashi}, \citenamefont {Andrae},
  \citenamefont {Balmaseda}, \citenamefont {Balsamo}, \citenamefont {Bauer},
  \citenamefont {Bechtold}, \citenamefont {{P., Beljaars}}, \citenamefont
  {van~de Berg}, \citenamefont {Bidlot}, \citenamefont {Bormann}, \citenamefont
  {C.}, \citenamefont {Dragani}, \citenamefont {Fuentes}, \citenamefont {Geer},
  \citenamefont {Haimberger}, \citenamefont {Healy}, \citenamefont {Hersbach},
  \citenamefont {H{\'{o}}lm}, \citenamefont {Isaksen}, \citenamefont
  {K{\aa}llberg}, \citenamefont {K{\"{o}}hler}, \citenamefont {Matricardi},
  \citenamefont {McNally}, \citenamefont {Monge-Sanz}, \citenamefont
  {Morcrette}, \citenamefont {Park}, \citenamefont {Peubey}, \citenamefont
  {de~Rosnay}, \citenamefont {Tavolato}, \citenamefont {Th{\'{e}}paut},\ and\
  \citenamefont {Vitart}}]{Dee2011}%
  \BibitemOpen
  \bibfield  {author} {\bibinfo {author} {\bibfnamefont {D.~P.}\ \bibnamefont
  {Dee}}, \bibinfo {author} {\bibfnamefont {S.~M.}\ \bibnamefont {Uppala}},
  \bibinfo {author} {\bibfnamefont {A.~J.}\ \bibnamefont {Simmons}}, \bibinfo
  {author} {\bibfnamefont {P.}~\bibnamefont {Berrisford}}, \bibinfo {author}
  {\bibfnamefont {P.}~\bibnamefont {Poli}}, \bibinfo {author} {\bibfnamefont
  {S.}~\bibnamefont {Kobayashi}}, \bibinfo {author} {\bibfnamefont
  {U.}~\bibnamefont {Andrae}}, \bibinfo {author} {\bibfnamefont {M.~A.}\
  \bibnamefont {Balmaseda}}, \bibinfo {author} {\bibfnamefont {G.}~\bibnamefont
  {Balsamo}}, \bibinfo {author} {\bibfnamefont {P.}~\bibnamefont {Bauer}},
  \bibinfo {author} {\bibnamefont {Bechtold}}, \bibinfo {author} {\bibfnamefont
  {A.~C.~M.}\ \bibnamefont {{P., Beljaars}}}, \bibinfo {author} {\bibfnamefont
  {L.}~\bibnamefont {van~de Berg}}, \bibinfo {author} {\bibfnamefont
  {J.}~\bibnamefont {Bidlot}}, \bibinfo {author} {\bibfnamefont
  {N.}~\bibnamefont {Bormann}}, \bibinfo {author} {\bibfnamefont
  {D.}~\bibnamefont {C.}}, \bibinfo {author} {\bibfnamefont {R.}~\bibnamefont
  {Dragani}}, \bibinfo {author} {\bibfnamefont {M.}~\bibnamefont {Fuentes}},
  \bibinfo {author} {\bibfnamefont {A.~J.}\ \bibnamefont {Geer}}, \bibinfo
  {author} {\bibfnamefont {L.}~\bibnamefont {Haimberger}}, \bibinfo {author}
  {\bibfnamefont {S.~B.}\ \bibnamefont {Healy}}, \bibinfo {author}
  {\bibfnamefont {H.}~\bibnamefont {Hersbach}}, \bibinfo {author}
  {\bibfnamefont {E.~V.}\ \bibnamefont {H{\'{o}}lm}}, \bibinfo {author}
  {\bibfnamefont {L.}~\bibnamefont {Isaksen}}, \bibinfo {author} {\bibfnamefont
  {P.}~\bibnamefont {K{\aa}llberg}}, \bibinfo {author} {\bibfnamefont
  {M.}~\bibnamefont {K{\"{o}}hler}}, \bibinfo {author} {\bibfnamefont
  {M.}~\bibnamefont {Matricardi}}, \bibinfo {author} {\bibfnamefont {A.~P.}\
  \bibnamefont {McNally}}, \bibinfo {author} {\bibfnamefont {B.~M.}\
  \bibnamefont {Monge-Sanz}}, \bibinfo {author} {\bibfnamefont {J.-J.}\
  \bibnamefont {Morcrette}}, \bibinfo {author} {\bibfnamefont {B.-K.}\
  \bibnamefont {Park}}, \bibinfo {author} {\bibfnamefont {C.}~\bibnamefont
  {Peubey}}, \bibinfo {author} {\bibfnamefont {P.}~\bibnamefont {de~Rosnay}},
  \bibinfo {author} {\bibfnamefont {C.}~\bibnamefont {Tavolato}}, \bibinfo
  {author} {\bibfnamefont {J.-N.}\ \bibnamefont {Th{\'{e}}paut}}, \ and\
  \bibinfo {author} {\bibfnamefont {F.}~\bibnamefont {Vitart}},\ }\href@noop {}
  {\bibfield  {journal} {\bibinfo  {journal} {Q.J.R. Meteorol. Soc}\ }\textbf
  {\bibinfo {volume} {137}},\ \bibinfo {pages} {553} (\bibinfo {year}
  {2011})}\BibitemShut {NoStop}%
\bibitem [{\citenamefont {Lenzen}\ \emph {et~al.}(2012)\citenamefont {Lenzen},
  \citenamefont {Kanemoto}, \citenamefont {Moran},\ and\ \citenamefont
  {Geschke}}]{Lenzen2012}%
  \BibitemOpen
  \bibfield  {author} {\bibinfo {author} {\bibfnamefont {M.}~\bibnamefont
  {Lenzen}}, \bibinfo {author} {\bibfnamefont {K.}~\bibnamefont {Kanemoto}},
  \bibinfo {author} {\bibfnamefont {D.}~\bibnamefont {Moran}}, \ and\ \bibinfo
  {author} {\bibfnamefont {A.}~\bibnamefont {Geschke}},\ }\href@noop {}
  {\bibfield  {journal} {\bibinfo  {journal} {Environ. Sci. Technol.}\ }\textbf
  {\bibinfo {volume} {46}},\ \bibinfo {pages} {8374} (\bibinfo {year}
  {2012})}\BibitemShut {NoStop}%
\bibitem [{\citenamefont
  {\url{https://confluence.ecmwf.int/display/CKB/ERA-Interim+issues+with+cloud+cover}}()}]{ERAINT}%
  \BibitemOpen
  \bibfield  {author} {\bibinfo {author} {\bibnamefont
  {\url{https://confluence.ecmwf.int/display/CKB/ERA-Interim+issues+with+cloud+cover}}},\
  }\href@noop {} {}\BibitemShut {NoStop}%
\bibitem [{\citenamefont {Keenlyside}\ and\ \citenamefont
  {Latif}(2007)}]{Keenlyside2007}%
  \BibitemOpen
  \bibfield  {author} {\bibinfo {author} {\bibfnamefont {N.~S.}\ \bibnamefont
  {Keenlyside}}\ and\ \bibinfo {author} {\bibfnamefont {M.}~\bibnamefont
  {Latif}},\ }\href@noop {} {\bibfield  {journal} {\bibinfo  {journal} {J.
  Clim.}\ }\textbf {\bibinfo {volume} {20}},\ \bibinfo {pages} {131} (\bibinfo
  {year} {2007})}\BibitemShut {NoStop}%
\bibitem [{\citenamefont {Barrat}\ \emph {et~al.}(2004)\citenamefont {Barrat},
  \citenamefont {Barthelemy}, \citenamefont {Pastor-Satorras},\ and\
  \citenamefont {Vespignani}}]{Vespignani2004}%
  \BibitemOpen
  \bibfield  {author} {\bibinfo {author} {\bibfnamefont {A.}~\bibnamefont
  {Barrat}}, \bibinfo {author} {\bibfnamefont {M.}~\bibnamefont {Barthelemy}},
  \bibinfo {author} {\bibfnamefont {R.}~\bibnamefont {Pastor-Satorras}}, \ and\
  \bibinfo {author} {\bibfnamefont {A.}~\bibnamefont {Vespignani}},\
  }\href@noop {} {\bibfield  {journal} {\bibinfo  {journal} {PNAS}\ }\textbf
  {\bibinfo {volume} {101}},\ \bibinfo {pages} {3747} (\bibinfo {year}
  {2004})}\BibitemShut {NoStop}%
\bibitem [{\citenamefont {Maluck}\ and\ \citenamefont
  {Donner}(2015)}]{Maluck2015}%
  \BibitemOpen
  \bibfield  {author} {\bibinfo {author} {\bibfnamefont {J.}~\bibnamefont
  {Maluck}}\ and\ \bibinfo {author} {\bibfnamefont {R.~V.}\ \bibnamefont
  {Donner}},\ }\href@noop {} {\bibfield  {journal} {\bibinfo  {journal} {PLoS
  One}\ }\textbf {\bibinfo {volume} {10}},\ \bibinfo {pages} {e0133310}
  (\bibinfo {year} {2015})}\BibitemShut {NoStop}%
\bibitem [{\citenamefont {Rheinwalt}\ \emph {et~al.}(2012)\citenamefont
  {Rheinwalt}, \citenamefont {Marwan}, \citenamefont {Kurths}, \citenamefont
  {Werner},\ and\ \citenamefont {Gerstengabe}}]{Rheinwalt2012}%
  \BibitemOpen
  \bibfield  {author} {\bibinfo {author} {\bibfnamefont {A.}~\bibnamefont
  {Rheinwalt}}, \bibinfo {author} {\bibfnamefont {N.}~\bibnamefont {Marwan}},
  \bibinfo {author} {\bibfnamefont {J.}~\bibnamefont {Kurths}}, \bibinfo
  {author} {\bibfnamefont {P.}~\bibnamefont {Werner}}, \ and\ \bibinfo {author}
  {\bibfnamefont {F.-W.}\ \bibnamefont {Gerstengabe}},\ }\href@noop {}
  {\bibfield  {journal} {\bibinfo  {journal} {EPL}\ }\textbf {\bibinfo {volume}
  {100}},\ \bibinfo {pages} {28002} (\bibinfo {year} {2012})}\BibitemShut
  {NoStop}%
\end{thebibliography}%

\end{document}